\documentclass[aip,reprint,amsmath,amssymb,unsortedaddress]{revtex4-2}

\usepackage[version=3]{mhchem} 
\usepackage[T1]{fontenc}       
\usepackage{miller}
\usepackage[utf8]{inputenc}
\usepackage{graphicx}
\usepackage{amsfonts}
\usepackage{epstopdf}
\usepackage{float}

\usepackage{gensymb}
\usepackage{datetime}
\usepackage{color}

\usepackage{soul}

\usepackage[english]{babel}

\usepackage[separate-uncertainty=true,detect-all,retain-explicit-plus]{siunitx}

\usepackage{bm}

\begin{document}

\title{Spotlight on Charge-Transfer Excitons in Crystalline Textured \textit{n}-Alkyl Anilino Squara\-ine Thin Films}

\author{Frank Balzer}
\affiliation{SDU Centre for Photonics Engineering, University of Southern Denmark, Alsion~2, DK-6400 S{\o}nderborg, Denmark.}

\author{Nicholas J. Hestand}
\affiliation{Department of Natural and Applied Sciences, Evangel University, Springfield, Missouri 65802, USA.}

\author{Jennifer Zablocki}
\affiliation{Kekulé-Institute for Organic Chemistry and Biochemistry, University of Bonn, Gerhard-Domagk-Str.~1, D-53121 Bonn, Germany.}

\author{Gregor Schnakenburg}
\affiliation{Institute of Inorganic Chemistry, University of Bonn, Gerhard-Domagk-Str.~1, D-53121 Bonn, Germany.}

\author{Arne Lützen}
\affiliation{Kekulé-Institute for Organic Chemistry and Biochemistry, University of Bonn, Gerhard-Domagk-Str.~1, D-53121 Bonn, Germany.}

\author{Manuela Schiek}
\email{manuela.schiek@jku.at}
\affiliation{LIOS \& ZONA, Johannes Kepler University Linz, Altenberger Str. 69, A-4040 Linz, Austria.}

\keywords{squaraines, polymorphism, Organic Molecular Beam Deposition (OMBD), Davydov-splitting, polarized light absorption}

\begin{abstract}
Prototypical \textit{n}-alkyl terminated anilino squaraines for photovoltaic applications show characteristic double-hump absorption features peaking in the green and deep-red spectral range. These signatures result from coupling of an intramolecular Frenkel exciton and an intermolecular charge transfer exciton. Crystalline, textured thin films suitable for polarized spectro-microscopy have been obtained for compounds with \textit{n}-hexyl (nHSQ) and \textit{n}-octyl (nOSQ) terminal alkyl chains. The here released triclinic crystal structure of nOSQ is similar to the known nHSQ crystal structure. Consequently, crystallites from both compounds show equal pronounced linear dichroism with two distinct polarization directions. The difference in polarization angle between the two absorbance maxima cannot be derived by spatial considerations from the crystal structure alone but requires theoretical modeling. Using an essential state model, the observed polarization behavior was discovered to depend on the relative contributions of the intramolecular Frenkel exciton and the intermolecular charge transfer exciton to the total transition dipole moment. For both nHSQ and nOSQ,
the contribution of the charge transfer exciton to the total transition dipole moment was found to be small compared to the intramolecular Frenkel exciton. Therefore, the net transition dipole moment is largely determined by the intramolecular component resulting in a relatively small mutual difference between the polarization angles. Ultimately, the molecular alignment within the micro-textured crystallites can be deduced and, with that, the excited state transitions can be spotted.
\end{abstract}

\maketitle

\section{Introduction}
Squaraine dyes are an attractive class of donor-acceptor-donor type molecular semiconductors that have been used in a variety of applications.\cite{He2020,Gsaenger2016,Halton08,Beverina10,Weiss16,Law93} Since their solid-state absorption and solute-state fluorescence are strong in the visible to near-infrared region, their purposes as p-type material range from solar cells\cite{Chen19,Maeda18,Chen13,Deing2012,Brueck14} and photosensors,\cite{Somashekharappa2020,Schulz19,Strassel2018,Binda11} even in biological environment, \cite{Abdullaeva18} to thin film transistors,\cite{Gsaenger2014} as well as photosensitizers in photodynamic therapy\cite{Avirah2012,Babu2017} and fluorescent labels.\cite{Sreejith2015,Wu2018}
Functionality typically arises from supramolecular interactions giving rise to new phenomena not present on a molecular level. Therefore, it is essential to understand molecular aggregation for the targeted design of opto-electronic and photonic applications. Aggregation both in colloidal solute systems and structured or extended thin films also desires comprehension from a theoretical perspective.  \cite{Saikin13,Painelli2004,Roesch2020,Shen2021,Shen2020,Roehr18,Bialas2021,Hestand2017}

\begin{figure}[h]
\centering
  \includegraphics[width=0.27\textwidth]{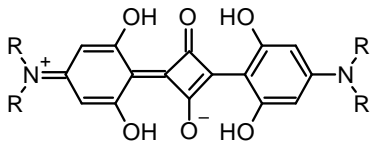}
  \caption{Schematic molecular structure of \textit{n}-alkyl anilino squaraines in a zwitterionic resonance formula, where R denotes the variable alkyl substituents. In this study, these are linear alkyl chain of 4, 5, 6, and 8 carbon atoms: \textit{n}-butyl (nBSQ), \textit{n}-pentyl (nPSQ), \textit{n}-hexyl (nHSQ), and \textit{n}-octyl (nOSQ).}
  \label{fgr:formula}
\end{figure}

Anilino squaraines (SQs), consisting of two anilino rings connected via a central squaric moiety, are of donor-acceptor-donor (DAD) type: The molecular backbone is coplanarized by intramolecular hydrogen bonding if four hydroxy groups attached to the anilino rings are adjacent to the central core, Figure~\ref{fgr:formula}. The degenerate zwitterionic resonance structures results in a quadrupolar character of the molecule and in a strong intramolecular charge transfer. This leads to intense intermolecular interactions involving excitonic coupling in the aggregated state. Terminal alkyl substituents are not part of the chromophoric system, however, they steer the molecular packing, and are thereby determinative of the excitonic coupling.

For example, bulky terminal functionalization leads to herringbone-like monoclinic and orthorhombic crystal phases.\cite{Balzer17a,Zablocki20} Since their primitive unit cells contain more than one molecule, the optical thin-film absorbance spectra are typically dominated by Davydov splitting, which can be sufficiently explained by a classical Frenkel exciton picture.\cite{Hestand18} In contrast, for linear \textit{n}-alkyl substitution triclinic crystal structures have been reported with only a single molecule per unit cell.\cite{Brueck14,Chen14,Dirk95,Hestand15,Zablocki20a} This excludes Davydov splitting as the reason for the absorbance spectra consisting of two broad spectral features. In this case short intermolecular molecular distances based on the slipped $\pi$-stacking allows for a pronounced intermolecular charge transfer (ICT). Based on theoretical modeling\cite{Hestand15,Zheng16} it has been shown that coupling between this ICT resonance and an intramolecular Frenkel exciton results in the double-humped absorbance spectra.\cite{Hestand15,Zheng16} Modeling of such spectra is generally based on the single crystal structure data. Therefore, this has been limited for the \textit{n}-alkyl SQs to compounds with an alkyl chain length of three (nPrSQ), four (nBSQ), and six (nHSQ) carbon atoms up to now.

Here, we provide the crystal structure of nOSQ, \textit{n}-octyl SQ with eight carbon atoms. Together with the recently published crystal structure of nPSQ,\cite{Zablocki20a} (\textit{n}-pentyl SQ, five carbon atoms) the impact of the ICT on the optical spectra can be quantified for the series nBSQ, nPSQ, nHSQ, and nOSQ. Upon spincoating, these compounds were shown to form extended pseudo-uniaxial thin films with the crystallographic \hkl{001} plane parallel to the substrate but effectively random in-plane orientation.\cite{Zablocki20a} Crystallites of sizes suitable for spectro-microscopy could also be obtained for
nHSQ and nOSQ via dropcasting or dipcoating, and these samples exhibited an interesting polarization dependence of the high and low energy peaks. A previously established essential state model that includes ICT\cite{Hestand15,Zheng16} is used to account for the distinct linear dichroism of textured, micro-crystalline samples. The calculated spectra reproduce well the measured polarized absorbance spectra of such ordered nHSQ and nOSQ thin film samples. The key to resolve the spatial polarization pattern with respect to the micro-morphology is to account for the relative contributions of the intramolecular Frenkel and intermolecular charge transfer excitons to the total transition dipole moment of the absorbing states.

\section{Experimental Section}

The \textit{n}-alkyl terminated anilino squaraines nHSQ and nOSQ have been synthesized via a catalyst-free condensation reaction as previously documented.\cite{Brueck14,Zablocki20a}

The single crystal structures of nBSQ, nPSQ (CCDC code 1987522) and nHSQ (CCDC code 962720) have already been published.\cite{Hestand15,Brueck14,Zablocki20a} Their crystallographic parameters are presented in Table~S1 in the Supporting Information together with the parameters of the newly determined nOSQ (CCDC code 2077835) single crystal structure. All unit cells adopt the space group \textit{P}-1. Single crystals of nOSQ were grown by vapor diffusion using dichloromethane as solvent and cyclohexane as anti-solvent within one month.\cite{Spingler2012}
The structure data of nOSQ were measured with a Bruker D8-Venture diffractometer at \SI{166}{K} using \ce{Cu}-K$\alpha$ radiation ($\lambda =\SI{1.54184}{\angstrom}$). For data analysis SHELXL version 2014/7 and OLEX2 inlcuding twinned data refinement have been used.\cite{Sheldrick08,Dolomanov09} Lattice parameters for nOSQ are triclinic, space group \textit{P}-1, $a=\SI{5.3556(4)}{\angstrom}$, $b=\SI{10.7619(6)}{\angstrom}$, $c=\SI{19.5626(12)}{\angstrom}$, $\alpha=\SI{86.503(5)}{\degree}$, $\beta=\SI{89.667(5)}{\degree}$, and $\gamma=\SI{78.726(5)}{\degree}$, $Z=1$.

Crystal dimensions \SI{0.600}{mm} $\times$ \SI{0.400}{mm} $\times$ \SI{0.200}{mm}, metallic bluish green plate, empirical formula \ce{C48 H76 N2 O2} and weight \SI{777.10}{amu}, $V=\SI{1103.68(12)}{\angstrom^3}$,  density \SI{1.169}{g/cm^3}, absorption coefficient \SI{0.592}{mm^{-1}}, F(000) = 426, multi-scan absorption correction (Bruker TWINABS-2012/1), 2$\Theta$ range for data collection \SI{4.526}{\degree} to \SI{129.988}{\degree}, index ranges $-6 \leq h \leq 6, -12 \leq k \leq 12, -22 \leq l \leq 22$, 7728 reflections collected, final $R$ indices ($I>2\sigma(I)$) $R1 = 0.1452$, $wR2 =  0.3551$, $R$ indices (all data) $R1 = 0.2371$, $wR2 = 0.0.3997$, full-matrix least-squares on F$^2$ refinement, GOF on F$^2$ = 1.158 for 7728 data and 0 restraint and 258 parameters, largest diff. peak and hole \SI{0.62}{e\angstrom^{-3}} and \SI{-0.50}{e\angstrom^{-3}}.

Micro-crystalline samples supported on objective slides (VWR float glass) have been obtained by dropcasting or dipcoating in case of nHSQ and nOSQ. The other compounds nBSQ and nPSQ showed less tendency to crystallization under these conditions and rather formed extended pseudo-uniaxial thin films.\cite{Zablocki20a} For dropcasting a few drops of a $\approx$ \SI{1}{mg/mL} solution of nHSQ or nOSQ in amylene stabilized chloroform were left to dry in ambient conditions for one to two hours. For dipcoating a glass slide was placed upright standing in a slim beaker filled with a $\approx$ \SI{0.1}{mg/mL} solution of nHSQ or nOSQ in amylene stabilized chloroform. The solvent was left to evaporate within one to two days in ambient conditions. The samples have been dried on a hotplate at \SI{100}{\celsius} under inert nitrogen-filled glove box conditions for 30 minutes.

X-ray diffraction (XRD) provided that for both compounds, nHSQ and nOSQ, but also for nBSQ and nPSQ the \hkl{001} face was parallel to the substrate.\cite{Zablocki20a} Views of the molecular arrangements along the \hkl[-100] direction and parallel to the \hkl(001) plane for nBSQ, nPSQ, nHSQ, and nOSQ are presented in Figure~S1, the top view onto the \hkl(001) planes for nHSQ and nOSQ in Figure~S2. The stacking direction for all molecules is the \hkl[100] direction. Polarized reflection microscope images of spincasted nHSQ and nOSQ films together with unpolarized absorbance spectra (beam diameter $\approx$ \SI{3}{mm}) are shown in Figure~S3 in the Supporting Information.

The morphology of the samples was determined by atomic force microscopy (AFM, JPK NanoWizard) in intermittent contact mode (Tap300-G BudgetSensors cantilevers). For AFM image analysis, Gwyddion has been used.\cite{Necas12}

Polarized optical microscopy is done by a Leica DMRME polarization microscope, either in reflection or in transmission. Spectral resolution is obtained by bandpass filters (Thorlabs FKB-VIS-10, FWHM \SI{10}{nm}) in the beam path. The sample is rotated by a computer controlled stage (Thorlabs PRM1Z8), whereas the directions of the linear polarizers in the microscope, either a single one or two crossed polarizers, are fixed. The polarization angle $\phi_{\mathrm{max}}^{r,t}$, for which the reflectivity (r) or the transmission (t) is largest, is found pixelwise via a discrete Fourier transform using ImageJ, see the Supporting Information.\cite{Thevenaz98,Schneider12,Bernchou09,Balzer13} For fiber-like crystallites, also the angle $\beta^{r,t}$ between the polarization angle $\phi_{\mathrm{max}}^{r,t}$ and the long fiber axis is determined in reflection and transmission.\cite{Rezakhaniha11,DeMay11,Schiek08,Balzer17} For spatially resolved polarized spectroscopy, a fiber-optics miniature spectrometer (Ocean Optics Maya2000) is coupled through a \SI{200}{\um} optical fiber to the camera port of the microscope. That way, light is collected from a spot of typically \SIrange{30}{70}{\um} in diameter, depending on the choice of the collection optics in front of the fiber and the magnification of the microscope objective.

\section{Modeling}
The absorbance spectra of the four compounds, nBSQ, nPSQ, nHSQ, and nOSQ are modeled using an extended version of Painelli's essential states model for DAD chromophores.\cite{Terenziani06,Sanyal16}
In the essential states model, each molecule is modeled as a collection of three essential states that correspond to the molecule's dominant resonance structures: the state $|\mathrm{N}\rangle$ describes the neutral \ce{DAD} structure, and the two degenerate states $|\mathrm{Z_1}\rangle$ and $|\mathrm{Z_2}\rangle$ describe the zwitterionic structures \ce{D^+A^-D} and \ce{DA^-D^+}, respectively.\cite{Terenziani06} The zwitterionic states lie higher in energy than the neutral state by $\eta_\mathrm{Z}$ and couple to the neutral state through $-t_\mathrm{Z}$.

In reference \citenum{Hestand15}, Painelli's model was extended to include ICT interactions to explain the double-hump absorption spectra observed for extended squaraine thin films. ICT between nearest neighbor molecules is incorporated by including the ionic states $|\mathrm{A}\rangle$ (\ce{DA^{-}D}), $|\mathrm{C}_1\rangle$ (\ce{D^{+}AD}), $|\mathrm{C}_2\rangle$ (\ce{DAD^{+}}) and $|\mathrm{Z}_3\rangle$ (\ce{D^{+}A^{-}D^{+}}). Charge transfer configurations of a dimer pair consisting of one molecule in the $|\mathrm{A}\rangle$ state and the other in either the $|\mathrm{C}_1\rangle$ or $|\mathrm{C}_2\rangle$ state are energetically offset from the state where both molecules are in the $|\mathrm{N}\rangle$ state by the ion pair energy, $\eta_\mathrm{CT}$. When one molecule in the $|\mathrm{A}\rangle$ state and the other is in the $|\mathrm{Z}_3\rangle$ state, the energy is offset by an additional amount $\eta_Z$. The neutral molecular states $|\mathrm{N}\rangle$, $|\mathrm{Z}_1\rangle$ and $|\mathrm{Z}_2\rangle$ couple to the ionic states $|\mathrm{A}\rangle$, $|\mathrm{C}_1\rangle$, $|\mathrm{C}_2\rangle$ and $|\mathrm{Z}_3\rangle$ through the ICT integral, $-t_\mathrm{CT}$, which accounts for the transfer of an electron between a donor on one molecule and the acceptor on its neighbor. Hence, the two parameters describing ICT are $-t_\mathrm{CT}$ and $\eta_\mathrm{CT}$.

Coulombic intermolecular interactions are included by allowing molecules to interact electrostatically when in zwitterionic or ionic states, and vibronic coupling is considered by treating each arm of the squaraine molecule as a harmonic oscillator whose equilibrium geometry depends on the electronic state of the molecule.
Complete details of the model Hamiltonian can be found in reference \citenum{Hestand15}.

To model the squaraine thin films, each system is treated as a dimer pair consisting of the nearest-neighbor $\pi$-stacked molecules along the crystalline $a$-axis. The geometry of the dimer system is taken directly from the crystal structure.\cite{Hestand15,Zablocki20a,Brueck14} Polarized and unpolarized absorbance spectra are calculated assuming that the incident light is normal to the \hkl(001) crystal plane. A polarization angle of \SI{0}{\degree} corresponds to a polarization vector that is parallel to the projection of the long molecular axis onto the \hkl(001) plane. Increasing polarization angles correspond to the polarizer being rotated in a counterclockwise direction.

With the exception of the ion pair energy, $\eta_\mathrm{CT}$, and the ICT integral, $t_\mathrm{CT}$, all parameters are the same as in reference \citenum{Hestand15}, see Table~S2 in the Supporting Information. Both $\eta_{\mathrm{CT}}$ and $t_\mathrm{CT}$ are expected to depend on packing geometry, and are therefore allowed to vary between the systems studied.
Specifically, $\eta_\mathrm{CT}$ and $t_\mathrm{CT}$ were optimized to reproduce the energy difference between the two main absorption peaks, $\Delta E = E_1-E_2$, and the ratio of their maximum intensity, $R = I_1/I_2$. The energy difference $\Delta E$ is sensitive mainly to $t_\mathrm{CT}$ while the peak ratio $R$ is sensitive mainly to $\eta_\mathrm{CT}$. The optimization was performed using the differential evolution algorithm in SciPy\cite{Virtanen20} with the objective of minimizing the sum of the squares of the residuals, ($\Delta E_\mathrm{sim} - \Delta E_\mathrm{exp})^2 + (R_\mathrm{sim} - R_\mathrm{exp})^2$. The remaining model parameters are all intramolecular parameters, and since the chromophore backbone is the same in all systems studied, they are not expected to vary significantly from system to system.

\section{Results and Discussion}

\subsection{Unpolarized Absorbance of Extended Thin Films}

\begin{figure}[t]
 \centering
 \includegraphics[width=0.45\textwidth]{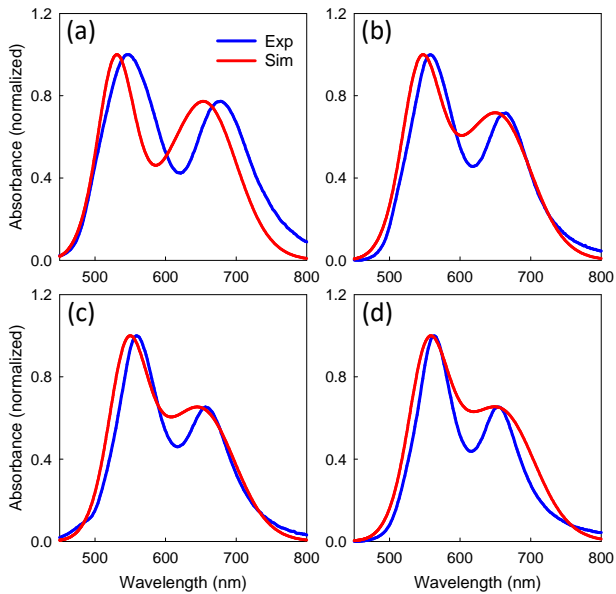}
 \caption{Simulated unpolarized absorbance spectra (red) compared to experiment (blue) for (a) nBSQ, (b) nPSQ, (c) nHSQ, and (d) nOSQ. Spetcra were measured on spincasted samples subjected to subsequently annealing at \SI{120}{\celsius}.\cite{Zablocki20a}}
 \label{fgr:H1}
\end{figure}

Figure~\ref{fgr:H1} shows the absorbance spectra of spincasted nBSQ, nPSQ, nHSQ, and nOSQ thin films. As documented previously,\cite{Zablocki20a} the absorbance spectra of these thin films do not depend on polarization because they are polycrystalline and the size of the observation spot for the spectrometer is at least an order of magnitude larger than the size of the randomly distributed crystallites. The spectra of all four compounds show two characteristic humps with a short- (high energy) and a long-wavelength (low energy) maximum roughly peaking at $\lambda \approx \SI{550}{nm}$ ($\approx$ \SI{2.25}{eV}) and $\lambda \approx \SI{650}{nm}$ ($\approx$ \SI{1.91}{eV}), respectively.

The essential states model was optimized to reproduce the difference in peak energy, $\Delta E$, and the peak height ratio, $R$, of the spincasted samples for each compound by varying $\eta_\mathrm{CT}$ and $t_\mathrm{CT}$, see Table~\ref{tbl:optparameters}. Note that both $t_\mathrm{CT}$ and $\eta_\mathrm{CT}$ decrease with increasing alkyl chain length. The decrease in $\eta_\mathrm{CT}$ with increasing chain length contradicts the trend in this parameter for nBSQ and nHSQ in previous publications.\cite{Hestand15,Zheng16} In those papers, $\eta_\mathrm{CT}$ increased from nBSQ to nHSQ, while here it decreased.
\footnote{We also note that $t_\mathrm{CT}$ for nBSQ and nHSQ are different than the values reported in \cite{Hestand15,Zheng16}. In those papers, $t_\mathrm{CT} = \SI{2742}{cm^{-1}}$ (\SI{0.34}{eV}) and \SI{2420}{cm^{-1}} (\SI{0.30}{eV}) for nBSQ and nHSQ, respectively. Note that in reference \citenum{Hestand15} the value for $t_\mathrm{CT}$ reported in Table~2 is incorrect. The value used in the simulations of nBSQ (called \ce{DBSQ(OH)2} in reference \citenum{Hestand15}) was \SI{0.34}{eV}, not \SI{0.28}{eV}.}

\begin{table}[h]
  \caption{Optimized modeling parameters: $\eta_\mathrm{CT}$ (mainly affects the peak ratio $R$) and $t_\mathrm{CT}$ (mainly affects the peak energy difference $\Delta E$).}
  \label{tbl:optparameters}
  \begin{tabular}{lrrrr}
    \hline
    Parameter & nBSQ & nPSQ & nHSQ & nOSQ  \\
    \hline
    $\eta_\mathrm{CT}~(\si{eV})$ & 1.4770 & 1.4700 & 1.4630 & 1.4540  \\
    $t_\mathrm{CT}~(\si{eV})$ & 0.3976 & 0.3395 & 0.3238 & 0.3116  \\
    \hline
  \end{tabular}
\end{table}

These differences could arise from different methods used to determine optimal $\eta_\mathrm{CT}$ and $t_\mathrm{CT}$, or different experimental spectra used to fit the parameters. In any case, given the semiempirical nature of the model, the parameters reported should be considered as effective parameters that are capable of illustrating trends when comparing different systems, not as highly accurate estimations of the true values. The overall trends in the parameters from the literature are consistent with the values found here.

The simulated spectra are compared to the experimental spectra in Figure~\ref{fgr:H1}. For all systems, the peak spacing, $\Delta E$, and ratio, $R$, of the simulated spectra are in good agreement with experiment. Especially the trend of decreasing $t_\mathrm{CT}$ and $\eta_\mathrm{CT}$ with increasing alkyl chain length clearly reproduces the relative increase of the short wavelength peak and narrowing of peak spacing, respectively, within the measured spectra, Figure~\ref{fgr:H1}.
The absolute peak energies, $E$, are in reasonable agreement with experiment, but the simulated spectra have slightly higher peak energies in all cases. The disagreement in the energies is not too surprising given the relative simplicity of the model. Compared to the real systems, for example, the simulated spectra only consider a dimer, not the full crystal. Other disagreements between the simulations and experiments include the peak broadening. In the simulated spectra, the high energy peak appears to be slightly narrower than the low energy peak, which is opposite the behavior observed in the experimental spectra (see also Figure~\ref{fgr:H5}c of reference \citenum{Zablocki20a}).

\subsection{Polarized Absorbance and Reflection of Microcrystalline Aggregates}

\begin{figure}[t]
 \centering
 \includegraphics[width=0.45\textwidth]{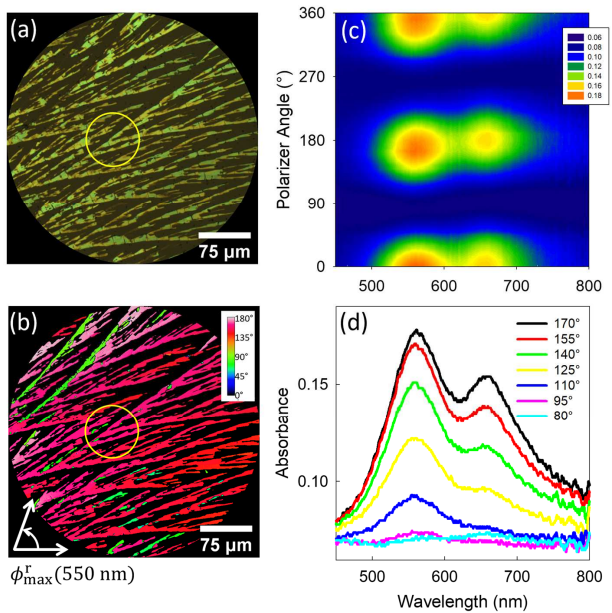}
 \caption{Polarized reflection and absorbance of nHSQ fibers on glass produced by dipcoating. A single polarizer reflection microscope image (a) demonstrates the formation of fiber-like structures. Analyzing the reflectivity at \SI{550}{nm} (b) shows that within the measurement area the angle for maximum reflectivity $\phi_{\max}^r(\SI{550}{nm})$ is well defined, i.e. the in-plane orientation of the molecules is almost parallel. The coordinate system for the angle is sketched in the lower left. Polarized absorbance spectra, (c) and (d), depict the polarization angle dependence of the two absorption peaks. The yellow circle in (a) and (b) marks the area from where absorbance spectra have been taken. The single spectra in (d) cover an polarization angle range over \SI{90}{\degree} in steps of \SI{15}{\degree}, starting at the short-wavelength maximum. The difference in polarizer angle between the short- and long-wavelength maximum is $\Delta \phi_{\max}^{\mathrm{spec}}=\SI{10\pm 3}{\degree}$.}
 \label{fgr:nhsq_polar}
\end{figure}

\begin{figure}[h]
 \centering
 \includegraphics[width=0.48\textwidth]{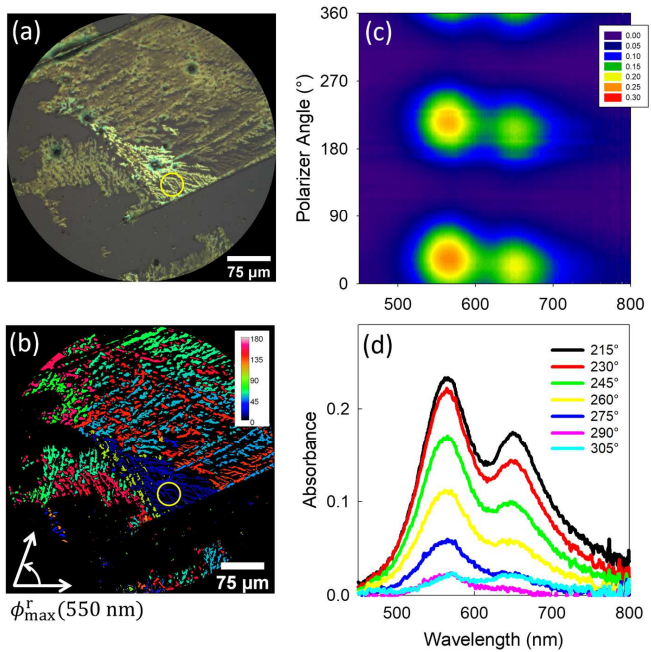}
 \caption{The same as Figure~\ref{fgr:nhsq_polar}, but for nOSQ aggregates produced by dipcoating. Note that for the selected nOSQ sample section for spectroscopy as indicated by the yellow circle in (a) and (b) the larger wavelength maximum appears at a smaller polarizer angle than the short wavelength maximum, opposite to the case of nHSQ. This time, for nOSQ the difference in polarizer angle between the short- and long-wavelength maximum obtained from transmission spectroscopy is $|\Delta \phi_{\max}^{\mathrm{spec}}|=\SI{11\pm 3}{\degree}$.}
 \label{fgr:nosq_polar}
\end{figure}

To get a better insight into the polarization dependence of the absorbance spectra, thin films with larger but isolated aggregates are advantageous. Here, such aggregates have been prepared by dipcoating and dropcasting. Large scale optical microscope images are presented in Figures~\ref{fgr:nhsq_polar}(a) and \ref{fgr:nosq_polar}(a), and in the Supporting Information section, where such aggregates can be clearly observed. AFM images, Figure~S4, demonstrate that the aggregates are several hundred nanometer tall and well isolated from each other.

The polarization analysis primarily targets the relative polarization difference of the two characteristic spectral features, specifically the relative difference in azimuthal polarization direction, $\Delta \phi$, of both peaks within the plane of the sample. The sample was eucentrically adjusted and rotated in steps of \SI{5}{\degree} while the linear polarizer was kept in a fixed position. The polarization analysis was conducted in two different ways: Pixelwise analysis of a series of microscopy images taken in reflection, as well as analysis of local transmission spectra. This returns the azimuthal polarization direction in reflection $\phi_{\max}^r$ and in transmission $\phi_{\max}^t$ from image analysis, and in transmission $\phi_{\max}^{\mathrm{spec}}$ from spectroscopic analysis, and from this also the relative differences. For suitable fiber-like aggregates the relative polarization angle between the long fiber axis and the polarization directions of each peak, $\beta^t$, is also determined from transmission image analysis.

For nHSQ, such a polarization analysis has been conducted for fiber-like aggregates near to the short and long-wavelength absorption maxima at $\lambda=\SI{550}{nm}$ (\SI{2.25}{eV}) and $\lambda=\SI{650}{nm}$ (\SI{1.91}{eV}), respectively. The reflectivity is also peaking at these spectral positions, and the polarization angles are well defined within microscopic regions. The angle for maximum reflectivity $\phi_{\max}^{r}(\SI{550}{nm})$ is shown in Figure~\ref{fgr:nhsq_polar}(b) while the angle for maximum reflectivity $\phi_{\max}^{r}(\SI{650}{nm})$ can be found in Figure~S5, Supporting Information. The difference $\Delta\phi_{\mathrm{max}}^r = \phi_{\mathrm{max}}^r(\SI{650}{nm}) - \phi_{\mathrm{max}}^r(\SI{550}{nm})$ has been calculated from the images as shown in Figure~S5(c) together with the corresponding histogramm in Figure~S5(d). This statistical analysis provides a positive and a negative maximum number, both having a value of $|\Delta\phi_{\mathrm{max}}^r|=\SI{7(3)}{\degree}$. The reason for both, a positive and a negative polarization angle difference, is two aggregates with mirrored contact planes that are \hkl(001) and \hkl(00-1) (summarized as equivalent planes \hkl{001}). For the chosen sample a negative $\Delta\phi_{\mathrm{max}}^r$ is more frequent, Figure~S5(c), and the respective fiber-like aggregates are color-coded red in Figure~\ref{fgr:nhsq_polar}(b). Adding a second, crossed polarizer to the illumination arm and detecting the polarizer angle for light extinction at the two reflection maxima leads to comparable results confirming their significance. This bireflection analysis is shown in the Supporting Information in Figures~S5(e) and (f).

The difference in polarization angle has also been determined from polarized transmission spectra obtained from the nHSQ sample section marked by a yellow circle in Figures~\ref{fgr:nhsq_polar}(a) and (b). From the transmission the absorbance $= -\log (T)$ is calculated, which is plotted in Figures~\ref{fgr:nhsq_polar}(c) and (d). The contour plot in (c) contains all spectra for a full turn of the linear polarizer, while (d) shows only selected spectra covering an angular range of \SI{90}{\degree} in steps of \SI{15}{\degree}, starting at the short-wavelength maximum.
The difference in polarization angle of the transmission spectroscopy maxima $\Delta\phi_{\mathrm{max}}^{\mathrm{spec}} = \phi_{\mathrm{max}}^{\mathrm{spec}}(\SI{650}{nm}) - \phi_{\mathrm{max}}^{\mathrm{spec}}(\SI{550}{nm})$ amounts to a slightly larger value of $\Delta\phi_{\mathrm{max}}^{\mathrm{spec}} = \SI{10\pm 3}{\degree}$.

To make sure, that this discrepancy between $\Delta\phi_{\mathrm{max}}^r$ and $\Delta\phi_{\mathrm{max}}^{\mathrm{spec}}$ is significant and not processing and morphology related, the polarization analysis has been repeated. This time, fiber-like nHSQ aggregates obtained from dropcasting have been investigated by single polarizer reflection and transmission imaging, see Supporting Information Figure~S8, and not by spectroscopic means. The difference in polarizer angle for the reflection measurement was reproduced being $|\Delta\phi_{\mathrm{max}}^r|=\SI{7(3)}{\degree}$ also for the dropcasted fiber-like nHSQ aggregates. Similarly, the difference in polarizer angle was found to be slightly larger for the transmission measurement amounting to $|\Delta\phi_{\mathrm{max}}^t|=\SI{9(3)}{\degree}$.

With that, the discrepancy between reflection and transmission analysis must be for a principal reason, which is related to the nature of crystallographic system and the amount of polarization rotation on its way through the sample. Whereas for orthorhombic systems the principal axes of both the index ellipsoid and the absorption ellipsoid have to agree with the crystal axes,\cite{Funke21,Bay20,Bay20a} this is not the case for triclinic crystals.\cite{Ramachandran61,Tompkins05} Triclinic unit cells are non-orthogonal systems while the index and absorption ellipsoids are always orthogonal systems. With that, their orientation is even wavelength dependent (axial dispersion).\cite{Dressel2008,Sturm2020} This also means that the polarizer angles for maxima and minima in reflected and transmitted intensity, respectively. Even under normal incidence they are not necessarily congruent for triclinic crystals. For nHSQ and nOSQ the index ellipsoid and the absorbance ellipsoid are essentially not known up to now. However, we estimate their orientational deviation from each other to be small since the inspected aggregate thickness is well below \SI{1}{\micro m}.

\begin{table}[h]
  \caption{\ Difference in polarization angle $\Delta\phi$ obtained from reflection and transmission images and spectroscopic transmission analysis as well as their average value.}
  \label{tbl:summarypolangles}
  \begin{tabular}{ccccc}
    \hline
    Compound & $|\Delta\phi_{\mathrm{max}}^r|$ (\si{\degree}) & $|\Delta\phi_{\mathrm{max}}^t|$ (\si{\degree}) & $|\Delta\phi_{\mathrm{max}}^{\mathrm{spec}}|$ (\si{\degree}) &  $|\Delta\phi_{\mathrm{max}}^{\mathrm{ave}}|$ (\si{\degree}) \\
    \hline
    nHSQ & \num{7(3)} & \num{9(3)} & \num{10(3)} & \num{9(5)} \\
    nOSQ & \num{8(3)} & \num{12(3)} & \num{11(3)} & \num{10(5)}\\
    \hline
  \end{tabular}
\end{table}

The same type of analysis has been performed on nOSQ micro-crystallites obtained from dipcoating, Figures~\ref{fgr:nosq_polar}(a)-(d). The nOSQ did not grow into such distinct fiber-like aggregates but rather showed fractal-like micro-aggregates with extended areas of homogeneous polarization. Here, the difference angle in transmission for the two absorbance maxima determined by spectroscopy is $|\Delta\phi_{\mathrm{max}}^{\mathrm{spec}}| =  \SI{11\pm 3}{\degree}$. Note that for the selected nOSQ sample section as indicated by the yellow circle in Figures~\ref{fgr:nosq_polar}(a) and (b) the larger wavelength maximum appears at a smaller polarizer angle than the short wavelength maximum, opposite to the case of nHSQ. Also for nOSQ the two mirror-imaged but otherwise equivalent crystallographic orientations \hkl(00-1) and \hkl(001) are realized.
The polarization analysis from imaging in transmission and reflection of the nOSQ sample is shown in the Supporting Information in Figure~S8. From the histograms in Figure~S8 (c) and (d), values of $|\Delta\phi_{\mathrm{max}}^t|=\SI{12(3)}{\degree}$ and $|\Delta\phi_{\mathrm{max}}^r|= \SI{8(3)}{\degree}$ have been determined, respectively. Again, the value for the difference in polarization angle is consistently larger for the transmission analysis of nOSQ for the reason of adopting a low-symmetry triclinic crystallographic unit cell.\cite{Ramachandran61,Tompkins05,Dressel2008,Sturm2020} In Table~\ref{tbl:summarypolangles} the experimentally determined values of all $\Delta\phi$ for nHSQ and nOSQ are summarized together with their calculated average. The respective average value $|\phi_{\mathrm{max}}^{\mathrm{ave}}|$ is used in the following for comparison with the simulated data.

\subsection{Origin of Polarization Dependence -- Modeling}

\begin{figure}[t]
 \centering
 \includegraphics[width=0.45\textwidth]{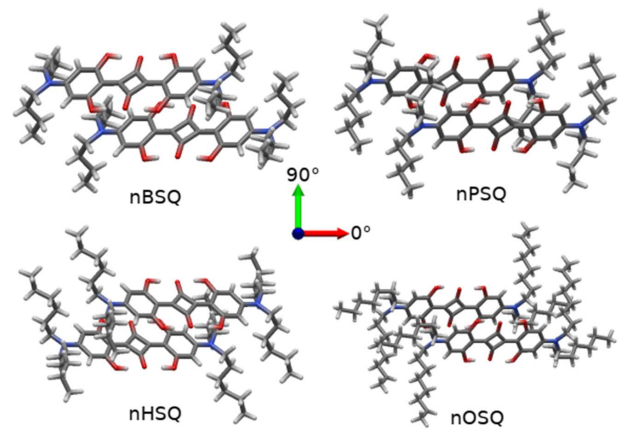}
 \caption{Dimer geometries in the orthogonal lab reference frame extracted from the single crystal structure data. Incident light travels along the $-z$ lab axis. Light is polarized at \SI{0}{\degree} when the polarization vector is parallel to the lab $x$-axis (red axis). The polarization angle increases in the counter clockwise direction so that light is polarized at \SI{90}{\degree} when the polarization vector is parallel to the lab $y$-axis (green axis). Note that for nHSQ the molecules $\pi$-stack ''up and to the right'' while for the other systems the molecules stack ''up and to the left''.}
 \label{fgr:H2}
\end{figure}

Polarized absorbance spectra were simulated using the parameters in Table~\ref{tbl:optparameters} and Supporting Information Table~S2. In all cases, the incident light is normal to the \hkl(001) crystal face. A polarization angle of \SI{0}{\degree} corresponds to a polarization vector that is parallel to the projection of the long molecular axis onto the \hkl(001) plane. As the polarization angle increases from \SI{0}{\degree}, the polarizer is rotated in a counterclockwise direction, see Figure~\ref{fgr:H2}.

\begin{figure}[t]
 \centering
 \includegraphics[width=0.45\textwidth]{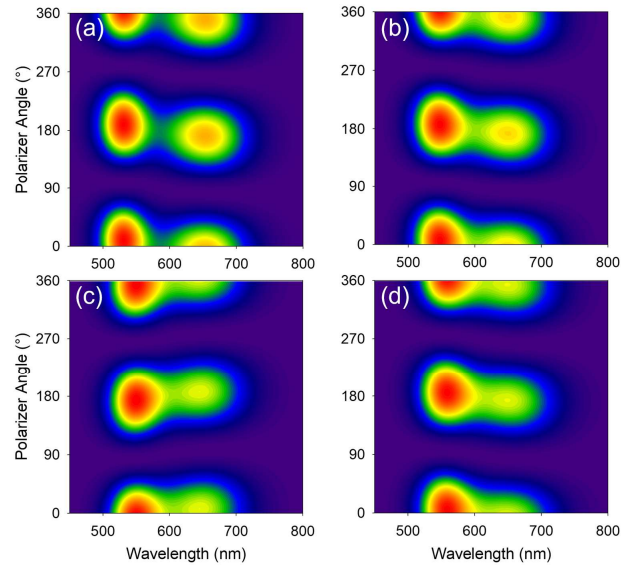}
 \caption{Simulated absorbance spectra for (a) nBSQ, (b) nPSQ, (c) nHSQ, and (d) nOSQ as a function of polarizer angle. Dark blue indicates regions of low absorbance and red indicates regions of high absorbance. Note that for nHSQ, the simulations predict that the short wavelength peak has a maximum at a smaller polarization angle than the long wavelength peak. This is a result of the different stacking geometry, see Figure~\ref{fgr:H2}, resulting in the CT dipole ''pointing'' in a different direction for nHSQ than the other systems. The simulations show the same behavior as the experiments, see Figures~\ref{fgr:nhsq_polar} and \ref{fgr:nosq_polar}.}
 \label{fgr:H3}
\end{figure}

Contour plots of the absorbance spectra as a function of polarizer angle are shown in Figure~\ref{fgr:H3}. For nHSQ and nOSQ, these can be compared to the experimental measurements shown in Figures~\ref{fgr:nhsq_polar} and \ref{fgr:nosq_polar}. The experimentally measured polarization dependence is reproduced well by the simulations. In particular, the low and high energy peaks in the simulated spectra show maxima at polarization angles separated by about $|\Delta\phi| = \SI{13}{\degree}$ and $|\Delta\phi| = \SI{12}{\degree}$ for nHSQ and nOSQ, respectively, see Table~\ref{tbl:polangles}. This is in good agreement with the experimental measurements for nHSQ and nOSQ, for which the peaks have maxima at angular differences of about $|\Delta \phi_{\mathrm{max}}^{\mathrm{ave}}| \approx \SI{10}{\degree}$.

\begin{figure}[t]
 \centering
 \includegraphics[width=0.45\textwidth]{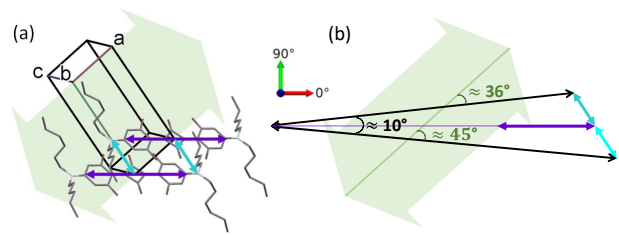}
 \caption{(a) A qualitative picture of the molecular (lilac) and intermolecular charge transfer (cyan) transition dipole moments for nHSQ, overlaid on the dimer in the orthogonal lab reference frame (red arrow \SI{0}{\degree}, green arrow \SI{90}{\degree}. The corresponding crystallographic unit cell projected onto the \hkl(00-1) plane is also shown. The broad light green arrow points out the possible direction of the long fiber-like aggregate axis, that is the crystallographic $a$-axis or \hkl[100]. (b) The black arrows represent a qualitative picture of the sum and difference of the molecular (lilac) and CT transition (cyan) dipole moments. Since the CT transition dipole moment is relatively small compared to the molecular, the directions of the net dipoles are largely determined by the molecular transition dipole moment. There is a relatively small angle (approx. $|\Delta \phi =  \SI{10}{\degree}|$) between the two measurable net transition dipole vectors (black arrows). These two net transition dipole moments have angles of about \SI{36}{\degree} and \SI{45}{\degree}, respectively, with the possible long fiber-like aggregate axis (broad light green arrow) being along the crystallographic $a$-axis.}
 \label{fgr:H5}
\end{figure}

The polarization dependence arises due to the different orientations of the molecular and charge transfer transition dipole moments, and their relative contribution to the total transition dipole moment of the absorbing states. The molecular transition dipole is oriented along the long axis of the molecule, while the charge transfer transition dipole is oriented along the $\pi$-stacking axis, see Figure~\ref{fgr:H5}(a). When the transition dipole moment of the absorbing state is dominated by a large molecular transition dipole component, absorbance is polarized mostly along the long molecular axis. In contrast, when the transition dipole moment is dominated by a large charge transfer transition dipole component, absorbance is polarized mostly along the $\pi$-stacking axis. By varying the molecular versus charge transfer composition of the transition dipole moment, the absorption polarization varies between these two extremes.

\begin{table}[h]
  \caption{\ Polarizer angle of maximum intensity for the high (peak 1) and low (peak 2) energy peaks for the simulated spectra. The angle difference $\Delta\phi$ between maxima is also reported. The experimentally determined value is the average over transmission and reflection measurements $|\Delta \phi_{\mathrm{max}}^{\mathrm{ave}}|$ as also given in Table~\ref{tbl:summarypolangles}.}
  \label{tbl:polangles}
  \begin{tabular}{ccccc}
    \hline
    Species & Peak 1 (\si{\degree}) & Peak 2 (\si{\degree}) & Calcd. $\Delta\phi$ (\si{\degree}) &  $\Delta \phi_{\max}^{\mathrm{ave}}$ (\si{\degree})\\
    \hline
    nBSQ & 189 & 172 & 17 & --  \\
    nPSQ & 187 & 173 & 14 & -- \\
    nHSQ & 174 & 187 & 13 & \num{9(5)} \\
    nOSQ & 186 & 174 & 12 & \num{10(5)}\\
    \hline
  \end{tabular}
\end{table}

To understand the origin of the polarization dependence, the transition dipole moment for each excited state was decomposed into its molecular (M) and charge transfer (CT) components, $\bm{\mu}_\mathrm{M}$ and $\bm{\mu}_\mathrm{CT}$, respectively, see Supporting Information for calculation details. Table~\ref{tbl:moments} shows the molecular and charge transfer components of the transition dipole moments for the states contributing the largest oscillator strength to each absorption peak.

For each state shown in Table~\ref{tbl:moments}, the charge transfer component, ${|\bm \mu}_\mathrm{CT}|$, is only about $10-20 \%$ as large as the molecular component, ${|\bm \mu}_\mathrm{M}|$. The large difference in magnitude between ${|\bm \mu}_\mathrm{CT}|$ and ${|\bm \mu}_\mathrm{M}|$ is the origin of the observed polarization dependence in Figure~\ref{fgr:H3}.

\begin{table*}[h]
  \caption{\ Total $|{\bm \mu}^{k\leftarrow 0}|$, molecular $|{\bm \mu}_\mathrm{M}^{k\leftarrow 0}|$, and charge transfer $|{\bm \mu}_\mathrm{CT}^{k\leftarrow 0}|$ ground to $k^{th}$ excited state transition dipole moments for the excited states of nBSQ, nPSQ, nHSQ, and nOSQ. Only the two states with the largest oscillator strengths are shown. Other states also contribute to the absorbance spectra. The values have been rounded to one decimal place.}
  \label{tbl:moments}
  \begin{tabular}{cccccc}
    \hline
    System & State $(k)$ & Energy (\si{eV}) & $|{\bm \mu}^{k\leftarrow 0}|$~(\si{D}) & $|{\bm \mu}_\mathrm{M}^{k\leftarrow 0}|$~(\si{D}) & $|{\bm \mu}_\mathrm{CT}^{k\leftarrow 0}|$~(\si{D}) \\
    \hline
    nBSQ & 105 & 1.8542 & 14.1 & 13.4 & 2.5 \\
         & 165 & 2.3455 & 11.5 & 11.0 & 1.9 \\ \\
    nPSQ & 108 & 1.8501 & 13.7 & 13.1 & 2.2 \\
         & 162 & 2.2818 & 15.2 & 14.7 & 2.1 \\ \\
    nHSQ & 108 & 1.8542 & 13.1 & 12.7 & 2.1 \\
         & 161 & 2.2731 & 15.6 & 15.1 & 2.0 \\ \\
    nOSQ & 107 & 1.8341 & 12.9 & 12.3 & 2.1 \\
         & 161 & 2.2418 & 15.5 & 15.1 & 1.9 \\
    \hline
  \end{tabular}
\end{table*}

\begin{table*}[h]
  \caption{\ The $x$ and $y$ components of the molecular and charge transfer transition dipole moments in the lab reference frame for each state listed in Table~\ref{tbl:moments}. The values have been rounded to one decimal place. See Figure~\ref{fgr:H2} for the definition of the lab reference frame.}
  \label{tbl:components}
  \begin{tabular}{cccccccc}
    \hline
    System & State $(k)$ & $\mu_x^{k\leftarrow 0}$~(\si{D}) & $\mu_y^{k\leftarrow 0}$~(\si{D}) & $\mu_{\mathrm{M},x}^{k\leftarrow 0}$~(\si{D}) & $\mu_{\mathrm{M},y}^{k\leftarrow 0}$~(\si{D}) & $\mu_{\mathrm{CT},x}^{k\leftarrow 0}$~(\si{D}) & $\mu_{\mathrm{M},y}^{k\leftarrow 0}$~(\si{D}) \\
    \hline
    nBSQ & 105 & -13.0 &  2.2 & -12.2 & 0.0 & -0.8 &  2.2 \\
         & 165 &  10.1 &  1.7 &  10.0 & 0.0 &  0.1 &  1.7 \\ \\
    nPSQ & 108 &  12.8 & -1.9 &  12.1 & 0.0 &  0.7 & -1.9 \\
         & 162 &  13.4 &  1.9 &  13.5 & 0.0 &  0.0 &  1.9 \\ \\
    nHSQ & 108 &  12.1 &  1.8 &  11.5 & 0.0 &  0.6 &  1.8 \\
         & 161 & -13.8 &  1.7 & -13.7 & 0.0 &  0.0 &  1.7 \\ \\
    nOSQ & 107 &  12.0 & -1.8 &  11.3 & 0.0 &  0.7 & -1.8 \\
         & 161 &  13.8 &  1.7 &  13.9 & 0.0 & -0.1 &  1.7 \\
    \hline
  \end{tabular}
\end{table*}

\begin{figure}[t]
    \centering
    \includegraphics[width=0.35\textwidth]{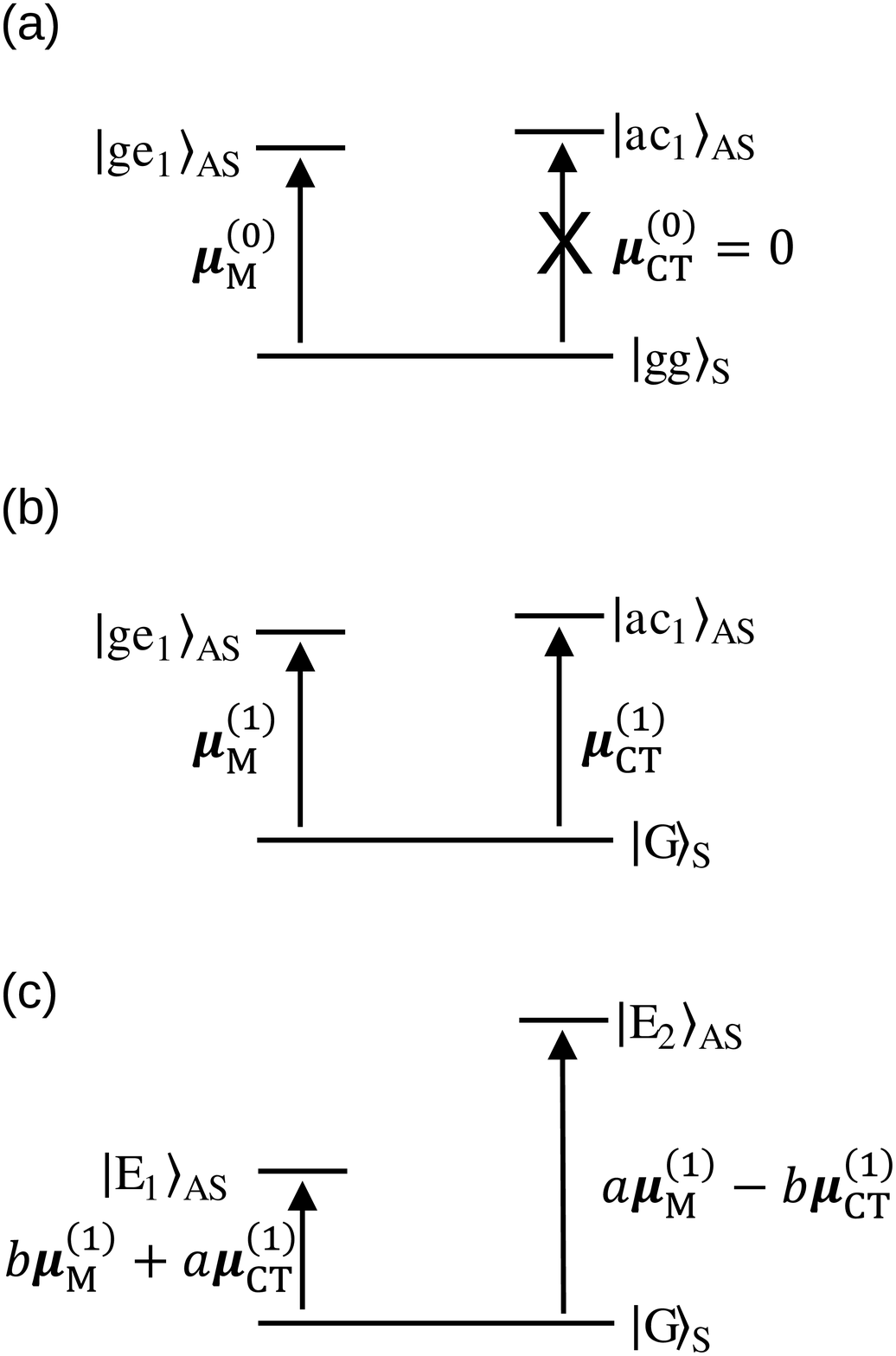}
    \caption{Energy level diagrams showing the electronic states relevant to the absorption processes in the squaraine systems considered here.
    (a) In the absence of intermolecular interactions and vibronic coupling, the ground state $|\mathrm{gg}\rangle_\mathrm{S}$ has zero charge transfer character, and only one excited state, $|\mathrm{ge}_{1}\rangle_\mathrm{AS}$, has a finite transition dipole moment, $\bm{\mu}_\mathrm{M}^{(0)}$.
    All other states, including $|\mathrm{ac}_{1}\rangle_\mathrm{AS}$ which is nearly resonant with $|\mathrm{ge}_{1}\rangle_\mathrm{AS}$, are dark.
    (b) When intermolecular charge transfer interactions are turned on, the ground state $|\mathrm{gg}\rangle_\mathrm{S}$ mixes with higher lying charge transfer states (not shown) to form a new ground state $|\mathrm{G}\rangle_\mathrm{S}$ with charge transfer character and a finite transition dipole to $|\mathrm{ac}_{1}\rangle_\mathrm{AS}$, $\bm{\mu}_\mathrm{CT}^{(1)}$.
    The new ground state also modifies the transition dipole to $|\mathrm{ge}_{1}\rangle_\mathrm{AS}$ to $\bm{\mu}_\mathrm{M}^{(1)}$.
    (c) Intermolecular charge transfer interactions cause mixing between $|\mathrm{ge}_{1}\rangle_\mathrm{AS}$ and $|\mathrm{ac}_{1}\rangle_\mathrm{AS}$ to create two new excited states $|\mathrm{E}_{1}\rangle_\mathrm{AS}\approx b|\mathrm{ge}_1\rangle_\mathrm{AS}+a|\mathrm{ac}_1\rangle_\mathrm{AS}$ and $|\mathrm{E}_{2}\rangle_\mathrm{AS}\approx a|\mathrm{ge}_1\rangle_\mathrm{AS}-b|\mathrm{ac}_1\rangle_\mathrm{AS}$ where $a$ and $b$ are the mixing coefficients.
    The transition dipole moments to these states are linear combinations of $\bm{\mu}_\mathrm{M}^{(1)}$ and $\bm{\mu}_\mathrm{CT}^{(1)}$ weighted by the coefficients $a$ and $b$.}
    \label{fgr:energydiagram}
\end{figure}

As discussed in reference\citenum{Hestand15}, the electronic states responsible for the absorbance spectra of squaraine thin films can be adequately described as a linear combination of the neutral and charge separated states $|\mathrm{ge}_1\rangle_\mathrm{AS}$ and $|\mathrm{ac}_1\rangle_\mathrm{AS}$.
The state $|\mathrm{ge}_1\rangle_\mathrm{AS}$ is the optically allowed intramolecular exciton while $|\mathrm{ac}_1\rangle_\mathrm{AS}$ is the lowest energy antisymmetric ICT state.
In the absence of intermolecular interactions and vibronic coupling, $|\mathrm{ge}_1\rangle_\mathrm{AS}$ and $|\mathrm{ac}_1\rangle_\mathrm{AS}$ are eigenstates of the system and $|\mathrm{ge}_1\rangle_\mathrm{AS}$ is the only state with a nonzero transition dipole moment, $\bm{\mu}_\mathrm{M}^{(0)}$, from the symmetric ground state $|\mathrm{gg}\rangle_\mathrm{S}$, see Figure~\ref{fgr:energydiagram}~(a).
In the absence of charge transfer interactions, $|\mathrm{gg}\rangle_\mathrm{S}$ has zero charge transfer character and absorbance to charge transfer states like $|\mathrm{ac}_{1}\rangle_\mathrm{AS}$ is forbidden.
Once the intermolecular charge transfer interactions are turned on, however, two important changes occur in the ground and excited state wave functions.
First, $|\mathrm{gg}\rangle_\mathrm{S}$ mixes with higher lying charge transfer states to form a new ground state $|\mathrm{G}\rangle_\mathrm{S}$ of mixed neutral and charge-transfer character, see Figure~\ref{fgr:energydiagram}~(b).
The new ground state results in new transition dipole moments from the ground state to $|\mathrm{ge}_1\rangle_\mathrm{AS}$ and $|\mathrm{ac}_1\rangle_\mathrm{AS}$, namely $\bm{\mu}_\mathrm{M}^{(1)}$ and $\bm{\mu}_\mathrm{CT}^{(1)}$.
Importantly, the transition to $|\mathrm{ac}_{1}\rangle_\mathrm{AS}$ is no longer forbidden, as the charge transfer character of $|\mathrm{G}\rangle_\mathrm{S}$ gives rise to a nonzero charge transfer transition dipole moment $\bm{\mu}_\mathrm{CT}^{(1)}$.
For the parameters relevant to the squarine systems studied here, the charge-transfer admixture to $|\mathrm{G}\rangle_{S}$ is small, which restricts $\bm{\mu}_\mathrm{CT}^{(1)}$ to be a small fraction of $\bm{\mu}_\mathrm{M}^{(1)}$, see Table~S3 in the Supporting Information.

\clearpage

The second important development that occurs when the charge transfer interactions are turned on is mixing of $|\mathrm{ge}_{1}\rangle_\mathrm{AS}$ and $|\mathrm{ac}_{1}\rangle_\mathrm{AS}$ to form new excited states $|\mathrm{E}_{1}\rangle_\mathrm{AS}\approx b|\mathrm{ge}_1\rangle_\mathrm{AS}+a|\mathrm{ac}_1\rangle_\mathrm{AS}$ and $|\mathrm{E}_{2}\rangle_\mathrm{AS}\approx a|\mathrm{ge}_1\rangle_\mathrm{AS}-b|\mathrm{ac}_1\rangle_\mathrm{AS}$ where $a$ and $b$ are the mixing coefficients, see Figure~\ref{fgr:energydiagram}~(c).
(We note that other higher-energy states mix into $|\mathrm{E}_{1}\rangle_\mathrm{AS}$ and $|\mathrm{E}_{2}\rangle_\mathrm{AS}$ to a smaller degree, but omit the details for the sake of clarity.)
For the parameters relevant to the squaraine systems studied here, $|\mathrm{ge}_{1}\rangle_\mathrm{AS}$ and $|\mathrm{ac}_{1}\rangle_\mathrm{AS}$ are near resonant so that $a$ and $b$ are of comparable magnitude.
The transition dipole moments to the new excited states are linear combinations of $\bm{\mu}_\mathrm{M}^{(1)}$ and $\bm{\mu}_\mathrm{CT}^{(1)}$ weighted by the coefficients $a$ and $b$.
For $|\mathrm{E}_{1}\rangle_\mathrm{AS}$ the dipole moment is approximately $b\bm{\mu}_\mathrm{M}^{(1)}+a\bm{\mu}_\mathrm{CT}^{(1)}$ and for $|E_{1}\rangle_\mathrm{AS}$ the dipole moment is approximately $a\bm{\mu}_\mathrm{M}^{(1)}-b\bm{\mu}_\mathrm{CT}^{(1)}$, see Figure~\ref{fgr:energydiagram}~(c).
Since $\bm{\mu}_\mathrm{M}^{(1)}$ and $\bm{\mu}_\mathrm{CT}^{(1)}$ have different orientations, absorbance to  $|\mathrm{E}_{1}\rangle_\mathrm{AS}$ and $|\mathrm{E}_{2}\rangle_\mathrm{AS}$ are polarized differently.
However, since the magnitude of $\bm{\mu}_\mathrm{CT}^{(1)}$ is small compared to $\bm{\mu}_\mathrm{M}^{(1)}$, and since $a$ and $b$ are of comparable magnitude, the difference in polarization angle between $|\mathrm{E}_{1}\rangle_\mathrm{AS}$ and $|\mathrm{E}_{2}\rangle_\mathrm{AS}$ is relatively small.

This behavior is demonstrated for the relevant states of the full simulations in Table~\ref{tbl:components}, which shows the $x$- and $y$-components of the total, molecular, and charge transfer dipole moments of the excited states in the lab reference frame. Since the CT component is small compared to the molecular component, the direction of the total transition dipole moment is determined mainly by the direction of the molecular component. Hence, the relatively small difference between the polarization angle of maximum intensity for the two absorption peaks.

A qualitative picture of this polarization behavior is shown for nHSQ in Figure~\ref{fgr:H5}.
Note that simple graphical vector addition of the lilac and cyan arrows indicating the molecular and CT transition dipole moments based on the projected structural parameters would not work. This approximation is only valid for excitons composed of transition dipole moments with equal oscillator strength and prevailing coulombic interactions.\cite{Balzer17a,Zablocki20}

For nHSQ fiber-like aggregates have been found, Figure~\ref{fgr:nhsq_polar} and Figures~S6 and S7. Therefore, the angle $\beta^\mathrm{t}$ of the polarized absorbance maximum with respect to the long fiber axis is well defined. The experimentally found angle $\beta_{\mathrm{550}}^\mathrm{t}$ at the short wavelength maximum is found to be $\SI{\pm 44}{\degree}\pm \SI{3}{\degree}$ and $\SI{\pm 45}{\degree}\pm \SI{3}{\degree}$ for the nHSQ samples presented in Figures~\ref{fgr:nhsq_polar} and S7, respectively. For the angle at the long wavelength maximum, $\beta_{\mathrm{650}}^\mathrm{t}$, two mirror imaged main values have been found for the two samples. These are $\SI{90}{\degree} - \SI{53(3)}{\degree} = \SI{37(3)}{\degree}$ and \SI{36(3)}{\degree}, respectively. Now considering that the growth direction of the long fiber axis is the molecular stacking direction, then the crystallographic $a$-axis or the \hkl[100] direction runs along the long nHSQ fiber axis. This in indicated by a broad light green arrow in Figure~\ref{fgr:H5}(a) overlaid on two nearest neighbor molecules and the crystallographic unit cell with \hkl(-001) orientation. The lilac and the cyan arrows indicate the directions of the molecular and the CT transition dipole moment, respectively, based on the sketched structural parameters. But since the CT transition dipole moment is comparatively small, the net transition dipole moments (black arrows) are largely determined by the molecular transition dipole moment, Figure~\ref{fgr:H5}(b). With an angle of approx. $|\Delta \phi =  \SI{10}{\degree}|$ in between these measurable excitonic net transition dipole moments, their projected angles onto the \hkl(-001) plane with respect to the long fiber axis (crystallographic $a$-axis) are about \SI{36}{\degree} and \SI{45}{\degree}, respectively, Figure~\ref{fgr:H5} (b). Indeed, this is observed by the experimental polarization analysis of the $\beta^\mathrm{t}$ angles supporting the picture of molecular orientation within the nHSQ fiber-like aggregates.

\section{Conclusions}
We have affirmed a previously established essential state model for the formation of hybrid intermolecular charge transfer and intramolecular Frenkel excitons for \textit{n}-alkyl terminated anilino squaraines in well ordered micro-crystalline textured thin films. The newly determined crystal structure of nOSQ (\textit{n}-octyl) extends the known series of nBSQ (\textit{n}-butyl), nPSQ (\textit{n}-pentyl) and nHSQ (\textit{n}-hexyl) and allows to point out clear trends for the intermolecular charge transfer integral $t_\mathrm{CT}$ and the ion pair energy $\eta_\mathrm{CT}$: both parameters decrease with increasing alkyl chain length. This is expressed in the absorbance spectra by a relative increase of the short wavelength spectral feature and a decrease in peak energy difference of both spectral features.
These generic double-hump spectral signatures show a characteristic linear dichroism with a relatively small angle between the two net transition dipole moment vectors, which cannot be understood by spatial considerations based on the crystallographic data of the triclinic unit cell. The key to resolve the spatial polarization pattern with respect to the micro-morphology is to account for the different contribution strengths of the molecular and charge transfer transition dipole moments. Quantum chemical modeling finds that the charge transfer transition dipole moment is small compared to the  molecular transition dipole moment, so that the directions of the measurable net transitions dipole moments is largely determined by the molecular component. This illustrates the power of spatially and polarization resolved analysis combined with theoretical considerations to obtain a deeper understanding of excitonic molecular states in micro-textured thin films.

\begin{acknowledgements}
MS thanks the PRO RETINA foundation (especially Franz Badura), as well as the Linz Institute of Technology (LIT-2019-7-INC-313 SEAMBIOF) for funding. JZ and AL gratefully acknowledges financial support from the DFG (RTG 2591 Template-designed Organic Electronics).
\end{acknowledgements}



\bibliography{Hestand_2022-Main}


\end{document}


\title{Supporting Information \\ Spotlight on Charge-Transfer Excitons in Crystalline Textured \textit{n}-Alkyl Anilino Squara\-ine Thin Films}

\author{Frank Balzer}
\affiliation{SDU Centre for Photonics Engineering, University of Southern Denmark, Alsion~2, DK-6400 S{\o}nderborg, Denmark.}

\author{Nicholas J. Hestand}
\affiliation{Department of Natural and Applied Sciences, Evangel University, Springfield, Missouri 65802, USA.}

\author{Jennifer Zablocki}
\affiliation{Kekulé-Institute for Organic Chemistry and Biochemistry, University of Bonn, Gerhard-Domagk-Str.~1, D-53121 Bonn, Germany.}

\author{Gregor Schnakenburg}
\affiliation{Institute of Inorganic Chemistry, University of Bonn, Gerhard-Domagk-Str.~1, D-53121 Bonn, Germany.}

\author{Arne Lützen}
\affiliation{Kekulé-Institute for Organic Chemistry and Biochemistry, University of Bonn, Gerhard-Domagk-Str.~1, D-53121 Bonn, Germany.}

\author{Manuela Schiek}
\email{manuela.schiek@jku.at}
\affiliation{LIOS \& ZONA, Johannes Kepler University, Altenberger Str. 69, A-4040 Linz, Austria.}

\makeatletter
\renewcommand{\thefigure}{S\@arabic\c@figure}
\renewcommand{\thetable}{S\@arabic\c@table}
\renewcommand{\theequation}{S\@arabic\c@equation}


\maketitle
\clearpage

\subsection{Crystal Structure}
\begin{figure}[h]
\centering
\includegraphics[width=0.7\textwidth]{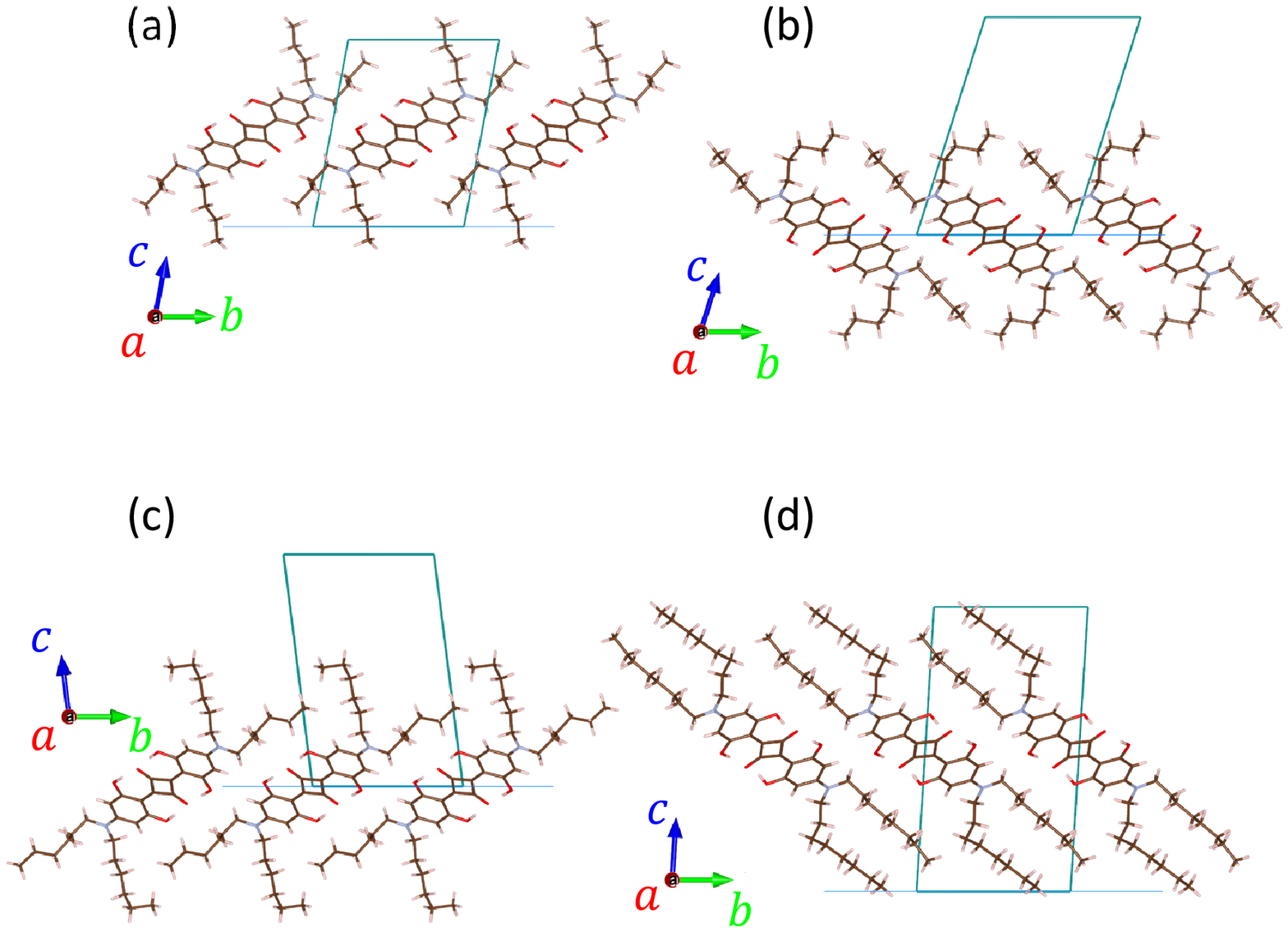}
\caption{Crystal structures of (a) nBSQ, (b) nPSQ, (c) nHSQ, and (b) nOSQ.}
\label{fgr:structure_extended}
\end{figure}

\begin{figure}[h]
\centering
\includegraphics[width=0.7\textwidth]{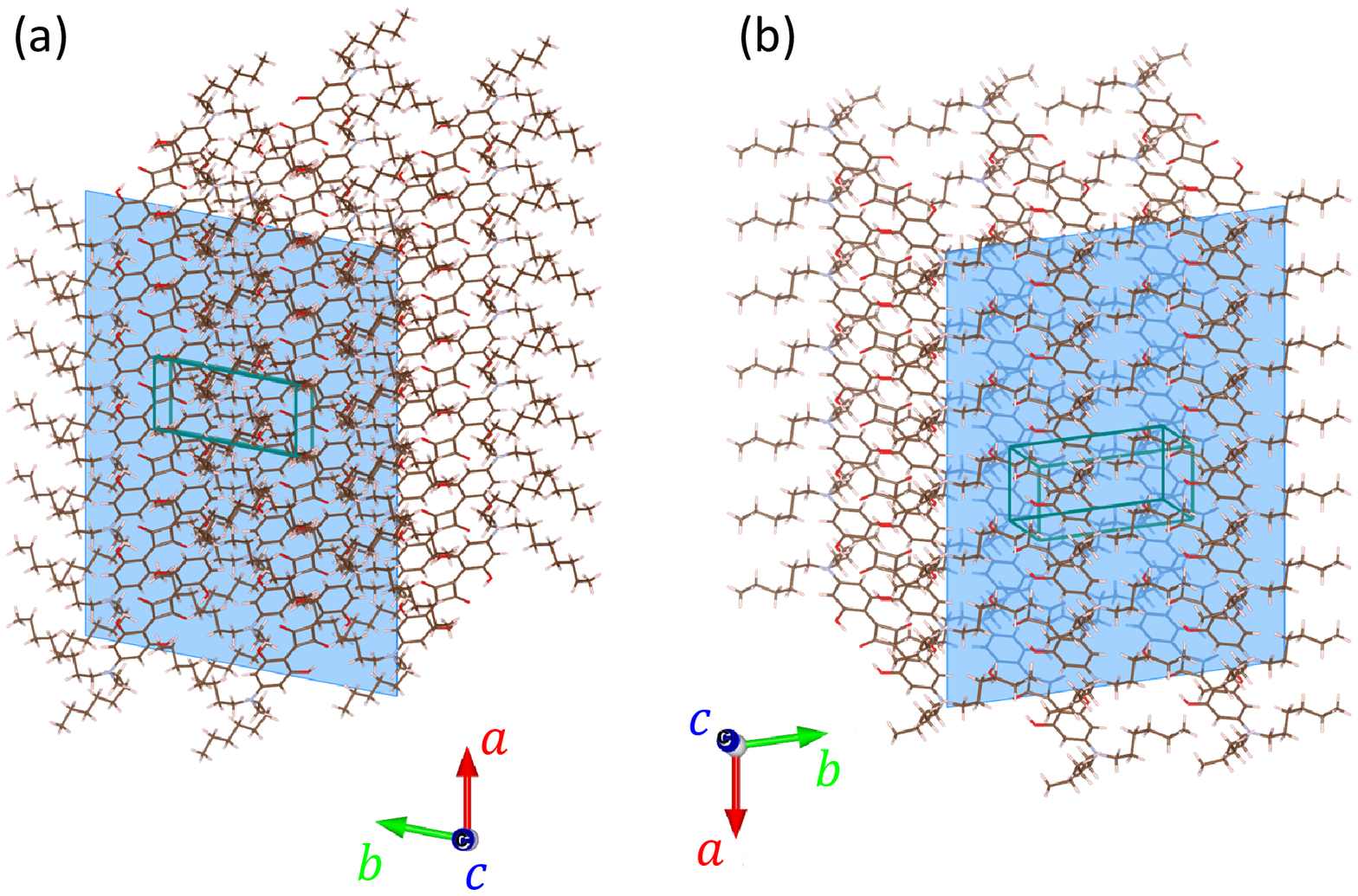}
\caption{Top view onto \hkl(001) planes of a single layer of (a) nHSQ and (b) nOSQ molecules. The \hkl(001) planes are sketched in blue.}
\label{fgr:structure_extended_001}
\end{figure}

A 3-dimensional view of the crystal structures of nBSQ,\cite{Hestand15} nPSQ,\cite{Zablocki20a} nHSQ,\cite{Brueck14} and nOSQ is presented in Figure~\ref{fgr:structure_extended}.\cite{Momma11} The trigonal unit cell axes are denoted by blue-green lines. The direction of view is along \hkl[-100], the horizontal plane is the \hkl(001) plane, which is parallel to the substrate surface. For nHSQ and nOSQ, Figure~\ref{fgr:structure_extended_001} shows the view onto the \hkl(001) plane.

\clearpage

\begin{figure}[t]
 \centering
 \includegraphics[width=0.5\textwidth]{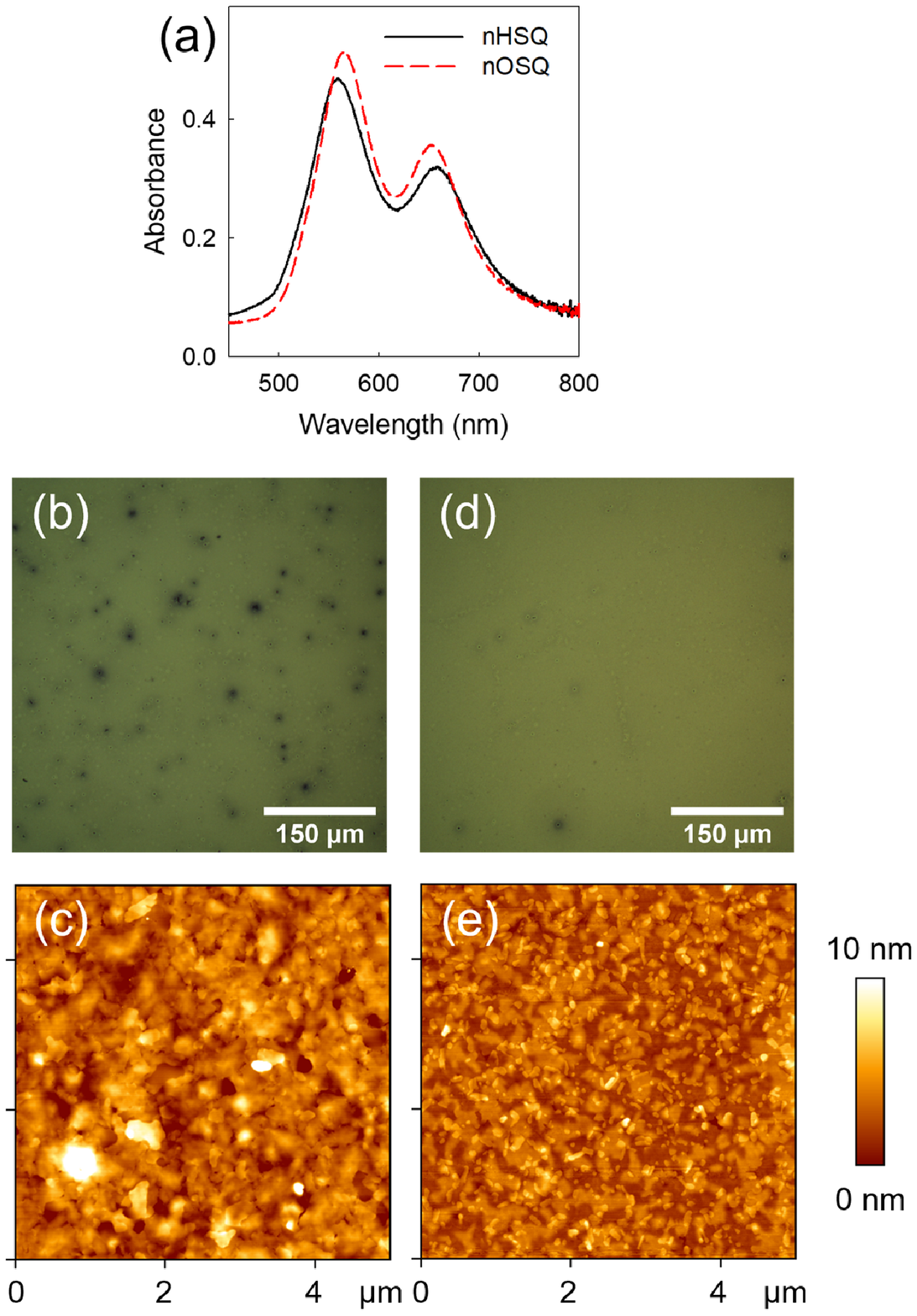}
 \caption{Unpolarized absorption spectra (Woollam M2000, beam diameter $\approx$ \SI{3}{mm}) in (a) of spincasted nHSQ and nOSQ thin films show two absorption maxima in the visible wavelength region. Polarized reflection microscope images (single horizontal polarizer) of nHSQ (b) and nOSQ (d) films on glass, respectively. Dark spots correspond to larger dumps of material. Corresponding AFM investigation of nHSQ (c) and nOSQ (e) reveals that the extended areas are rather flat. Their RMS roughness $S_q$ is generally below \SI{2}{nm}.}
 \label{fgr:spectra_spin}
\end{figure}

\begin{figure}[t]
 \centering
 \includegraphics[width=0.55\textwidth]{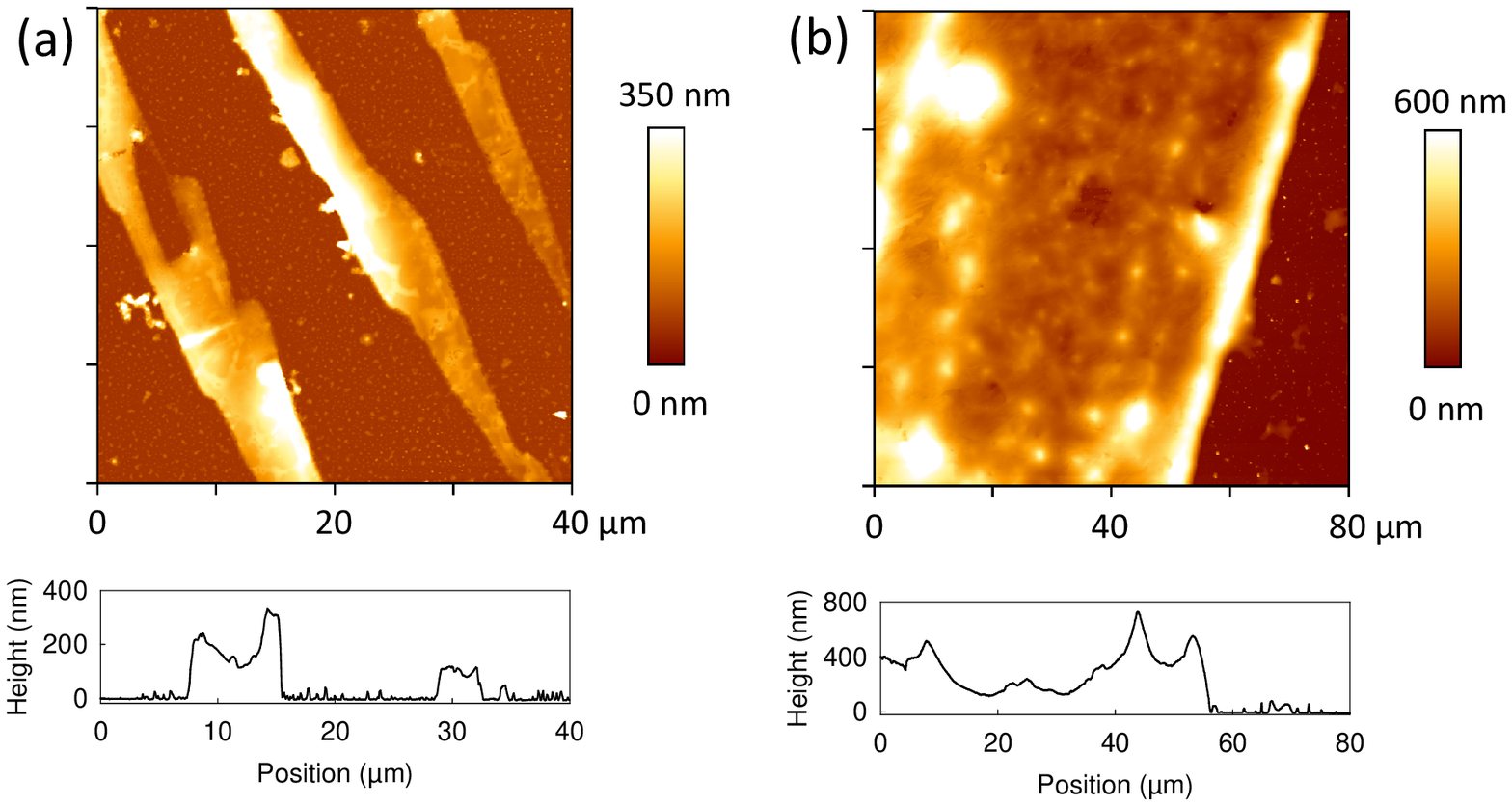}
 \caption{AFM images of (a) nHSQ fibers prepared by dipcoating and (b) nOSQ prepared by dropcasting. The cross-sections are along the \SI{10}{\micro m} lines.}
 \label{fgr:AFM_MS}
\end{figure}

\clearpage

\begin{table}[h]
    \caption{Single crystal structural data and calculated geometric parameters for some \textit{n}-alkyl anilino squaraines: CCDC code for structure; temperature ($T$) for single crystal measurement and unit cell parameters $a, b, c$ and $\alpha, \beta,\gamma$. The molecular arm length $L$ is half the distance between nitrogen atoms, the charge transfer length $ctL$ the shortest distance between nitrogen and the center of the squaric core between $\pi$-stacked molecules. The short axis slip $\Delta x$, long axis slip $\Delta y$, interplanar $\pi$-stacking distance $\Delta z$ are related to the translational distance $\Delta r$ between $\pi$-stacked molecules (equals value of $a$-axis). The angle between long molecule axis (LMA) and the $ab$-plane $\theta$-LMA and the angle between the direction of intermolecular charge transfer (ICT) and the $ab$-plane $\theta$-ctL are given, and finally the calculated $d$-spacing for \hkl(001) plane (Cu K$\alpha$-radiation). Columns for nBSQ, nPSQ, and nHSQ have been taken from \cite{Zablocki20a}. From the multiple data for nBSQ the structural parameters provided in reference\cite{Hestand15} were used for the modeling, because they have also been used within the previous modeling studies.\label{tab1}}
    \begin{tabular}{c|ccc|c|c|c}
  \hline
 compound &  \mbox{nBSQ\cite{Dirk95}} & \mbox{nBSQ\cite{Chen14}} & \mbox{nBSQ\cite{Hestand15}} & \mbox{nPSQ}\cite{Zablocki20a} & \mbox{nHSQ\cite{Brueck14}} & \mbox{nOSQ} \\
 CCDC &  1305855 & 958817 & -- & 1987522 & 962720 & 2077835 \\
 $T$ in K & 283-303 & 93 & 100 & 123 & 120 & 166 \\
 \hline
\mbox{a in \AA} &  5.260(1) & 5.226(1) & 5.169(4)  & 5.2359(7) & 5.097(2) & 5.3556(4)\\
\mbox{b in \AA} & 10.965(3) & 10.949(2) & 10.846(9)  & 10.7245(15) & 10.746(5) & 10.7619(6) \\
\mbox{c in \AA} & 13.786(4) & 13.701(2) & 13.538(11)  & 15.513(2) & 16.604(7) & 19.5626(12) \\
\hline
\mbox{$\alpha$ in $^\circ$} &  77.14(2) & 77.27(3) & 77.165(19) & 72.10(1)   & 96.374(2)  & 86.503(5) \\
\mbox{$\beta$ in $^\circ$}  &  80.08(2) & 79.54(1) & 79.143(14) & 85.856(11) & 94.825(11) & 89.667(5) \\
\mbox{$\gamma$ in $^\circ$} &  77.15(2) & 76.57(2) & 76.324(14) & 79.041(11) & 97.872(11) & 78.726(5) \\
\hline
\mbox{$L$ in \AA}   & 6.68 & 6.70 & 6.61 & 6.70 & 6.68 & 6.69\\
\mbox{$ctL$ in \AA} & 4.96 & 5.00 & 4.95 & 4.86 & 4.88 & 4.78 \\
\hline
\mbox{$\Delta x$ in \AA} & 1.86 & 1.93 & 1.91 & 1.70 & 1.66 & 1.76 \\
\mbox{$\Delta y$ in \AA} & 3.57 & 3.52 & 3.46 & 3.64 & 3.50 & 3.76 \\
\mbox{$\Delta z$ in \AA} & 3.39 & 3.34 & 3.32 & 3.35 & 3.31 & 3.38 \\
\hline
\mbox{$\Delta r$ in \AA \textsuperscript{(a)}} & 5.26 & 5.23 & 5.17 & 5.24 & 5.10 & 5.36 \\
\hline
\mbox{$\theta$-LMA in $^\circ$} & 23.8 & 23.7 & 23.7 & 23.4 & 24.4 & 23.5 \\
\mbox{$\theta$-ctL in $^\circ$} & 33.1 & 33.0 & 32.9 & 33.5 & 34.8 & 33.9 \\
\hline
\mbox{$d \hkl(001)$ in \AA} & 13.33 & 13.24 & 13.06 & 14.76 & 16.42 & 19.53 \\
\hline
 \end{tabular}

  (a) control: $\Delta r = \sqrt{\Delta x^2 + \Delta y^2 + \Delta z^2}$
\end{table}

\subsection{Modeling}
The extended essential states model is fully described in the main text and Supporting Information of reference \citenum{Hestand15}.
In that paper, the interested reader may find the complete Hamiltonian, equations used to calculate absorbance spectra, and expressions for the $|\mathrm{gg}\rangle_\mathrm{S}$, $|\mathrm{ge}_1\rangle_\mathrm{AS}$ and $|\mathrm{ac}_1\rangle_\mathrm{AS}$ states discussed in the main text of the current paper.
Here, we describe the formalism for calculating the molecular and charge transfer components of the dipole moment, provide the model parameters common to all systems studied, and present calculations of the charge transfer composition of the ground state for each system studied.

We begin by describing the formalism for calculating the molecular and charge transfer components of the dipole moment.
The total dipole moment operator $\hat{\mu}$ is defined as\cite{Hestand15}
\begin{equation}
 \hat{\mu} = \sum_{\mathrm{S}}\bm{\mu}(\mathrm{S})|\mathrm{S}\rangle\langle \mathrm{S}|
\end{equation}
where $\bm{\mu}(\mathrm{S})$ is the dipole moment of the electronic state $|\mathrm{S}\rangle$.
Calculation of $\bm{\mu}(\mathrm{S})$ is described in the Supporting Information of reference \citenum{Hestand15}.
For a dimer system, there are fifteen possible electronic states $|\mathrm{S}\rangle$ of the system, which may be divided into two subsets: one containing only neutral molecules
$S_\mathrm{M}=\{ \mathrm{NN}, \mathrm{NZ}_1, \mathrm{NZ}_2, \mathrm{Z}_1\mathrm{N}, \mathrm{Z}_1\mathrm{Z}_1, \mathrm{Z}_1\mathrm{Z}_2, \mathrm{Z}_2\mathrm{N}, \mathrm{Z}_2\mathrm{Z}_1, \mathrm{Z}_2\mathrm{Z}_2 \}$
and one containing only ionic molecules
$S_\mathrm{CT}=\{\mathrm{C}_1\mathrm{A}, \mathrm{C}_2\mathrm{A}, \mathrm{Z}_3\mathrm{A}, \mathrm{A} \mathrm{C}_1, \mathrm{A}C_2, \mathrm{AZ}_3\}.$
In the notation above, the first letter of the pair indicates the electronic state of molecule 1 and the second letter indicates the electronic state of molecule 2; for example, $\mathrm{NZ_{1}}$ indicates that molecule 1 is in the neutral state $|\mathrm{N}\rangle$ and molecule 2 is in the zwitterionic state $|\mathrm{Z_{1}}\rangle$.

Given these two subsets of electronic states, the total dipole operator may be decomposed into a molecular dipole operator
\begin{equation}
 \hat{\mu}_\mathrm{M} = \sum_{\mathrm{S}\in S_\mathrm{M}}\bm{\mu}(\mathrm{S})|\mathrm{S}\rangle\langle \mathrm{S}|
\end{equation}
where the sum runs over only the neutral molecular electronic states in $S_\mathrm{M}$, and an ionic dipole operator
\begin{equation}
 \hat{\mu}_\mathrm{CT} = \sum_{\mathrm{S}\in S_\mathrm{CT}}\bm{\mu}(\mathrm{S})|\mathrm{S}\rangle\langle \mathrm{S}|
\end{equation}
where the sum runs over only the ionic electronic states in $S_{\mathrm{CT}}$.
The total dipole operator is then given by
\begin{equation}
 \hat{\mu} = \hat{\mu}_\mathrm{M} + \hat{\mu}_\mathrm{CT}.
\end{equation}

The eigenstates of the system are found numerically by diagonalizing the Hamiltonian matrix\cite{Hestand15} and then arranged in order of increasing energy.
The ground state is denoted by $|\psi_{0}\rangle$ and the $k^{th}$ excited state is denoted by $|\psi_{k}\rangle$
The total ground-to-$k^{th}$-excited state transition dipole moment is determined by taking the expectation value of the dipole operator between the ground and excited state:
\begin{equation}
 \mathbf{\mu}^{k\leftarrow 0}=\langle\psi_k|\hat{\mu}|\psi_0\rangle
\end{equation}
The total transition dipole moment may then be decomposed into the molecular component $\mathbf{\mu}_\mathrm{M}^{k\leftarrow 0}$ and the charge transfer component $\mathbf{\mu}_\mathrm{CT}^{k\leftarrow 0}$
\begin{equation}
 \mathbf{\mu}^{k\leftarrow 0}=\langle\psi_k|\hat{\mu}|\psi_0\rangle=\langle\psi_k|\hat{\mu}_\mathrm{M}+\hat{\mu}_\mathrm{CT}|\psi_0\rangle=\langle\psi_k|\hat{\mu}_\mathrm{M}|\psi_0\rangle+\langle\psi_k|\hat{\mu}_\mathrm{CT}|\psi_0\rangle= \mathbf{\mu}_\mathrm{M}^{k\leftarrow 0}+ \mathbf{\mu}_\mathrm{CT}^{k\leftarrow 0}.
\end{equation}
These are the quantities reported in the main text.

Next, Table~\ref{tbl:parameters} reports the model parameters that are common to all systems considered in the main text. These parameters are taken directly from reference \citenum{Hestand15}.

\begin{table}[h]
  \caption{Parameters common to all squaraine systems studied in the current manuscript, reproduced from reference \citenum{Hestand15}. With the exception of the line broadening parameter $\sigma$, these are all intramolecular parameters. Since the squaraine backbone does not change between the systems studied, these parameters are not expected to vary significantly between the systems. Note that here the line width is reported as the standard deviation of the Gaussian broadening function, $\sigma$ while in reference \citenum{Hestand15} it was reported as $\sqrt{2}\sigma$.}
  \label{tbl:parameters}
  \begin{tabular}{lr}
    \hline
    Parameter & Value  \\
    \hline
    $\eta_\mathrm{Z}~(\si{cm^{-1}})$ & 5564  \\
    $t_\mathrm{Z}~(\si{cm^{-1}})$ & 8468  \\
    $\hbar \omega_\mathrm{vib}~(\si{cm^{-1}})$ & 1255  \\
    line width, $\sigma~(\si{cm^{-1}})$ & 850  \\
    $\lambda^2$ & 1 \\
    $\lambda^2_\mathrm{CT}$ & 0.5 \\
    $v_\mathrm{max}$ & 5 \\
    \hline
  \end{tabular}
\end{table}

Finally, the charge transfer character of the ground state is quantified for each system in Table~\ref{tbl:character}.
The charge transfer character of the ground state is determined from the expectation value
\begin{equation}
 \langle\psi_{0}|\sum_{\mathrm{S}\in S_\mathrm{CT}} |\mathrm{S}\rangle \langle \mathrm{S}|\psi_{0}\rangle.
\end{equation}
As discussed in the manuscript, the small charge transfer admixture in the ground state develops due to interactions with higher lying charge transfer states\cite{Hestand15} and is responsible for the charge transfer component of the transition dipole moment.
The data in Table~\ref{tbl:character} show that the charge transfer admixture ranges from 5.3\% in nBSQ to 3.6\% in nOSQ, which is consistent with the decrease in $t_\mathrm{CT}$ from nBSQ to nOSQ.
Moreover, it is interesting to note that as the charge transfer admixture decreases, the average ratio of $|\bm{\mu}_{CT}|$ to $|\bm{\mu}_{M}|$ also decreases, presumably due to the decreasing charge transfer component of the dipole moment.
This results in a simultaneous decreases in the calculated $\Delta \phi$ as the transition dipole moments become increasingly dominated by the molecular component of the transition dipole moment.

\begin{table}[h]
  \caption{The charge transfer character (CT) of the ground state for each squaraine system. A correlation between the ground state charge transfer character, the ratio of $|{\bm \mu}_\mathrm{CT}| / |{\bm \mu}_\mathrm{M}|$, and the polarization angle difference between peak maxima is observed. Here $\langle |{\bm \mu}_\mathrm{CT} | / |{\bm \mu}_\mathrm{M} |\rangle$ is the average $|{\bm \mu}_\mathrm{CT}| / |{\bm \mu}_\mathrm{M}|$ ratio for the states shown in Table~4 of the main text. }
  \label{tbl:character}
  \begin{tabular}{cccc}
    \hline
    System & CT (\% ) & $\langle |{\bm \mu}_\mathrm{CT}| / |{\bm \mu}_\mathrm{M}|\rangle$ & Calcd. $\Delta\phi$ (\si{\degree})  \\
    \hline
    nBSQ & 5.3 & 0.179 & 17  \\
    nPSQ & 4.1 & 0.153 & 14  \\
    nHSQ & 3.7 & 0.147 & 13  \\
    nOSQ & 3.6 & 0.148 & 12  \\
    \hline
  \end{tabular}
\end{table}

\clearpage

\subsection{Polarization Analysis}
The samples are illuminated with unpolarized light through a bandpass filter. The reflected or transmitted light passes a linear polarizer, finally entering either a camera (PixeLINK PL-B873-CU) or a fiber optics spectrometer (Ocean Optics Maya2000). The sample is rotated in steps of \SI{5}{\degree}, and from a series of microscope images the angles $\phi_{\max}^{r,t}$ for maximum reflection (r) or transmission (t) are extracted by a discrete Fourier transform for each image pixel. \cite{Balzer13,Balzer15} Image processing is performed by ImageJ.\cite{Schneider12,Thevenaz98,Bernchou09}

\begin{figure}[h]
\centering
\includegraphics[width=0.48\textwidth]{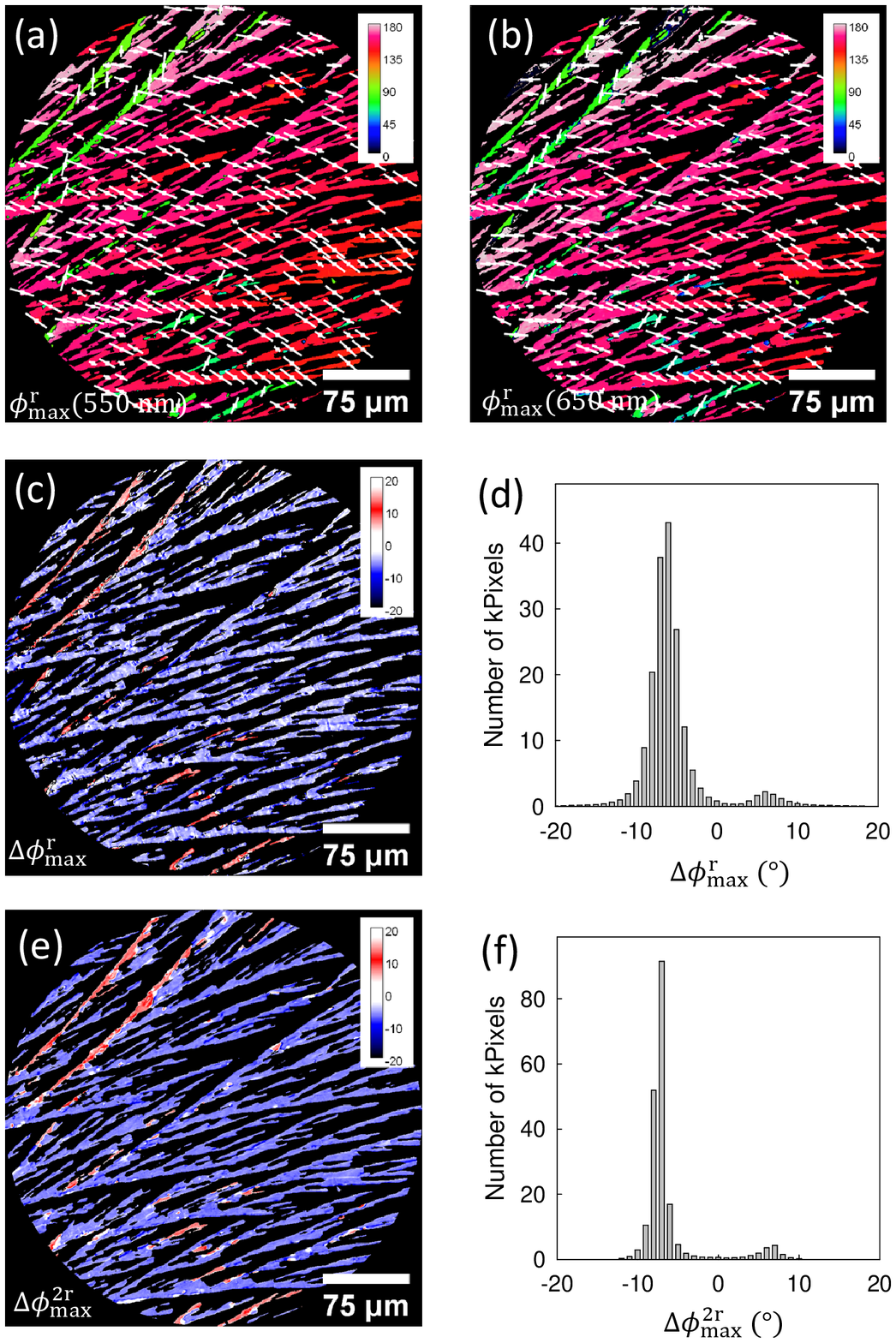}
\caption{Polarization analysis of nHSQ aggregates (single polarizer, reflection) at (a) $\lambda=\SI{550}{nm}$ and (b) \SI{650}{nm}. The polarizer angles $\phi_{\max}^r(\SI{550}{nm})$ and $\phi_{\max}^r(\SI{650}{nm})$, where reflectivity is largest, are encoded by colors and by short white lines. In (c) the difference between these two images is plotted, together with a histogram of the difference angle $\Delta\phi_{\max}^r$ in (d). Images (e) and (f) are the same as (c) and (d), but extracted from a polarization series using two crossed polarizers. In both cases the angles $|\Delta\phi_{\max}^r|$ and $|\Delta\phi_{\max}^{2r}|$ are \SI{7(3)}{\degree}.}
\label{fgr:n-HSQ_difference}
\end{figure}

The results for nHSQ are shown in Figure~\ref{fgr:n-HSQ_difference}. An analysis of polarized reflection images (single polarizer) at (a) $\lambda=\SI{550}{nm}$ and (b) $\lambda=\SI{650}{nm}$ reveals the polarizer directions, where reflectivity is largest, Figure~\ref{fgr:n-HSQ_difference}. The directions are encoded by colors, but also by short white lines. The difference between these two images, (c), reveals that there are two different fiber-like aggregates, where the difference angle is either positive or negative. This means, that either the \hkl(001) or the \hkl(00-1) face is parallel to the substrate. The histogram of the difference angle in (d) depicts that in reflection the average difference angle is $|\Delta\phi_{\max}^r|=\SI{7(3)}{\degree}$. This angle is somewhat smaller than the angle observed from polarized spectroscopy from the same spot with the same optical setup, Figure~3(c), probably because of a worse signal-to-noise ratio of the microscope camera compared to the spectrometer and since the images were taken in reflection, not in transmission.

Performing the same experiment with two crossed polarizers instead of a single one and determining the extinction angles for \SI{550}{nm} and \SI{650}{nm} leads to similar results, even so sharper since maxima occur every \SI{90}{\degree} instead of every \SI{180}{\degree}. The widths of the distributions are related to the experimental setup and not perfectly aligned images. Aggregates might also not to \SI{100}{\percent} be single crystalline. Error bars are estimated from the spread of several measurements, not from the width of the distributions.

\begin{figure}[h]
\centering
\includegraphics[width=0.5\textwidth]{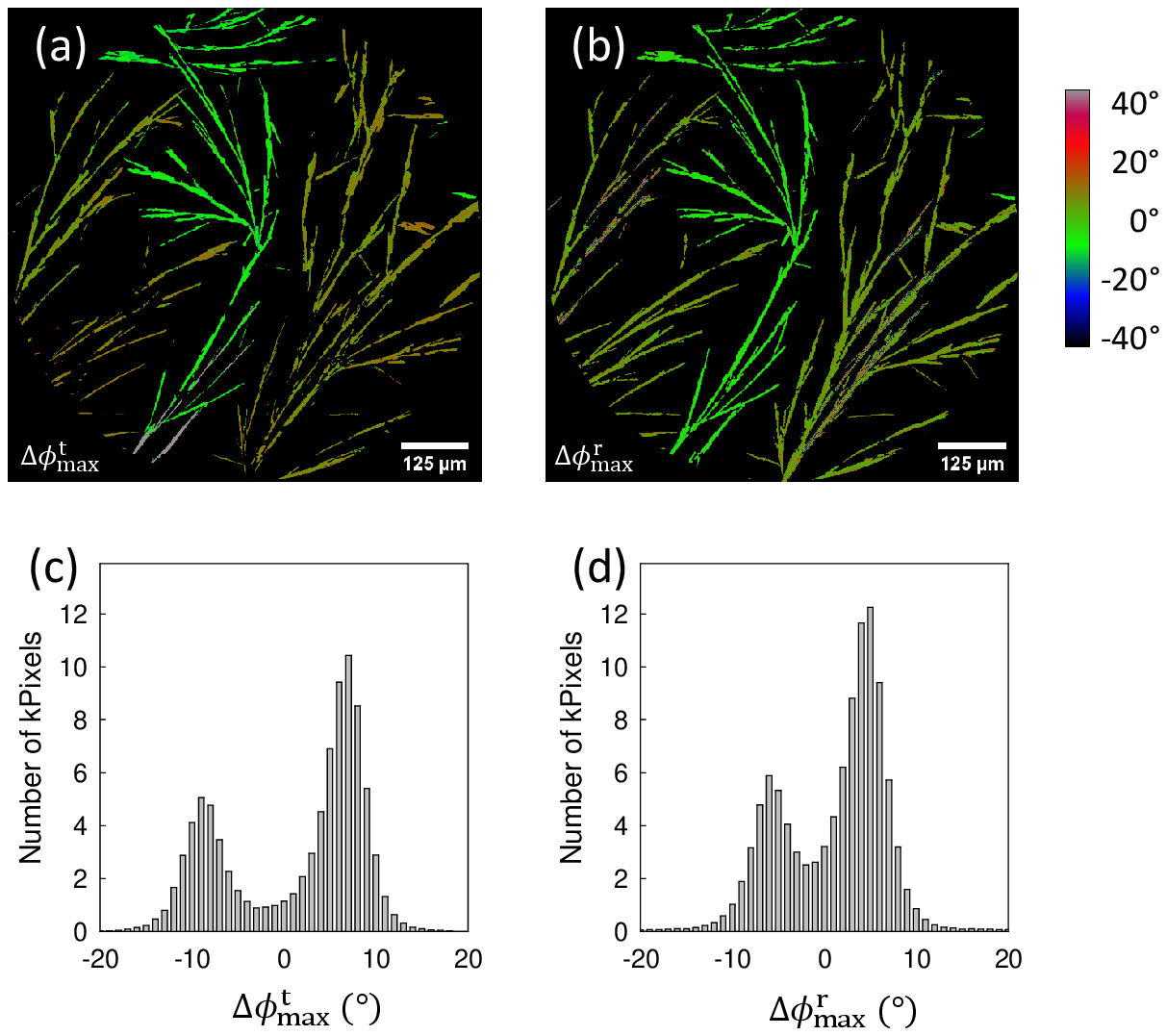}
\caption{Analysis of polarized transmission and reflection microscope images for nHSQ. For each set of measurements, the polarizer angle for either maximum reflectivity or maximum transmission has been determined pixelwise. Differences of polarizer angles taken with $\lambda=\SI{650}{nm}$ and $\lambda=\SI{550}{nm}$ for maximum transmission $\Delta\phi_{\max}^{t}$ (a), and for maximum reflection $\Delta\phi_{\max}^{r}$ (b), lead to peaks at either  $|\Delta\phi_{\max}^{t}|=\SI{9(3)}{\degree}$ (c) or $|\Delta\phi_{\max}^{r}|=\SI{7(3)}{\degree}$ (d).}
\label{fgr:n-HSQ_trans_refl}
\end{figure}

A similar analysis using transmitted light for a dropcasted nHSQ sample is shown in Figure~\ref{fgr:n-HSQ_trans_refl}. Here, the angles obtained are $|\Delta\phi_{\max}^t|=\SI{9(3)}{\degree}$ and $|\Delta\phi_{\max}^r|=\SI{7(3)}{\degree}$. Since the resulting nHSQ crystallites are fiber-like, it is easier to determine the angles $\beta^\mathrm{t}$ and $\beta^{\mathrm{r}}$ between the maximum transmission or reflection and the long fiber axis.

\clearpage

\begin{figure}[h]
\centering
\includegraphics[width=0.45\textwidth]{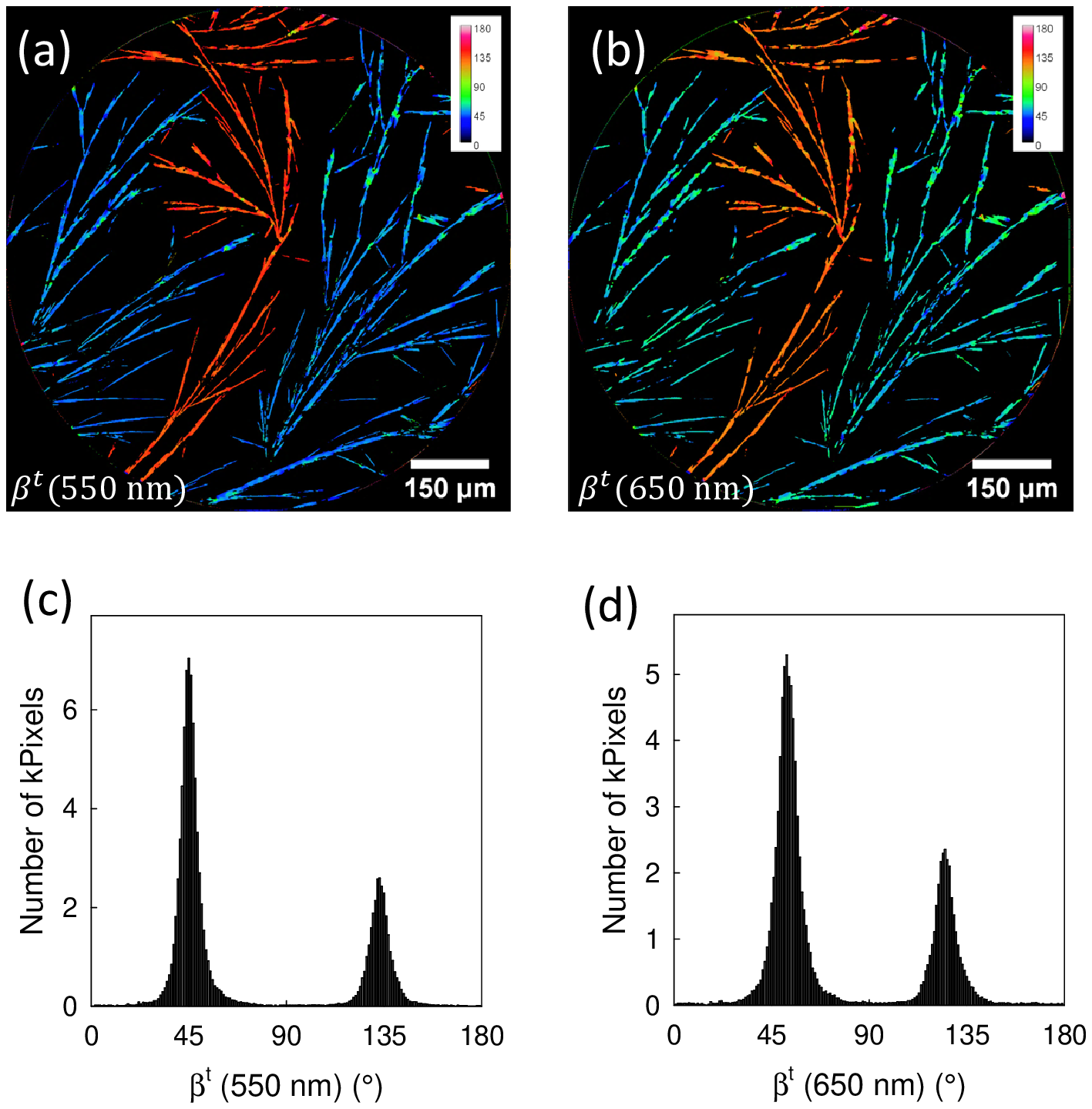}
\caption{Polarization analysis of nHSQ fibers. Mapping of the difference angle $\beta^\mathrm{t}$ between the long molecular axis and the maximum in transmission (single polarizer) at (a) $\lambda=\SI{550}{nm}$ and (b) $\lambda=\SI{650}{nm}$. The angles are encoded by colors. In (c) and (d) histograms of $\beta^\mathrm{t}$ at \SI{550}{nm} and \SI{650}{nm}, respectively, are shown, peaking at approximately \SI{90\pm 45}{\degree} and \SI{90\pm 35}{\degree}. }
\label{fgr:n-HSQ_beta}
\end{figure}

For transmission, these angles are color encoded in (a) and (b) in Figure~\ref{fgr:n-HSQ_beta} for $\lambda=\SI{550}{nm}$ and $\lambda=\SI{650}{nm}$. Corresponding histograms are shown in (c) and (d), respectively. The distributions peak at \SI{90}{\degree} $\pm$ \SI{45(4)}{\degree} for \SI{550}{nm}, and at \SI{90}{\degree} $\pm$ \SI{36(4)}{\degree} for \SI{650}{nm}.

\begin{figure}[h]
\centering
\includegraphics[width=0.45\textwidth]{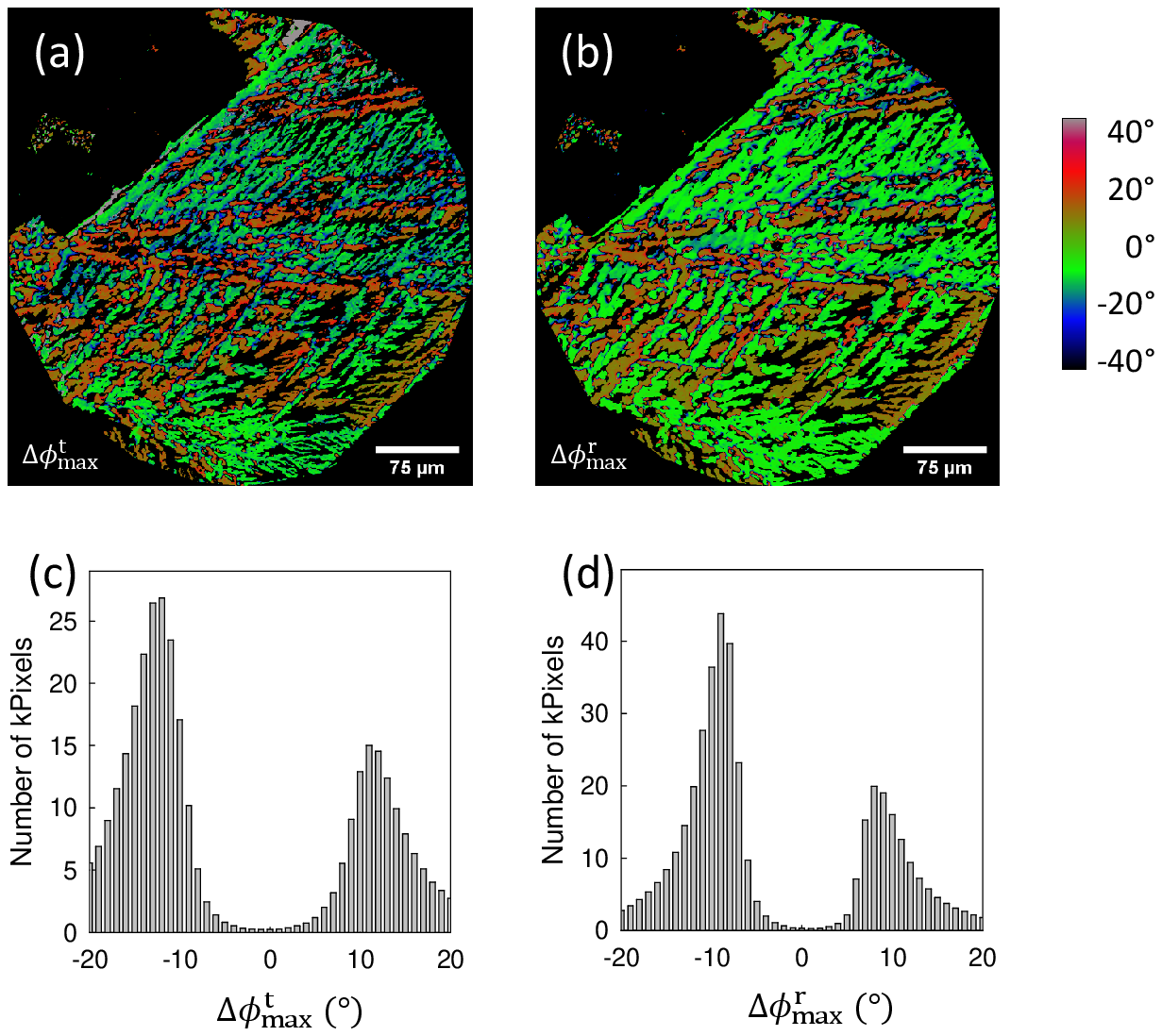}
\caption{Analysis of polarized transmission and reflection microscope images for nOSQ. For each set of measurements, the polarizer angle for either maximum reflectivity or maximum transmission has been determined pixelwise. Differences of polarizer angles taken with $\lambda=\SI{650}{nm}$ and $\lambda=\SI{550}{nm}$ for maximum transmission $\Delta\phi_{\max}^{t}$ (a), and for maximum reflection $\Delta\phi_{\max}^{r}$ (b), lead to peaks at either \SI{12(3)}{\degree} (c) or \SI{8(3)}{\degree} (d).}
\label{fgr:n-OSQ_trans_refl}
\end{figure}

For nOSQ, the obtained aggregates are more densely packed. Nevertheless, a polarization analysis is possible both in transmission and in reflection, Figure~\ref{fgr:n-OSQ_trans_refl}. Distributions of the difference angles $\Delta\phi_{\mathrm{max}}^\mathrm{t}$ and $\Delta\phi_{\mathrm{max}}^\mathrm{r}$, Figure~\ref{fgr:n-OSQ_trans_refl}(c) and (d) are broader, but still show pronounced peaks at \SI{12(3)}{\degree} and \SI{8(3)}{\degree}, respectively.

\clearpage

\bibliography{Hestand_2022-SI}